\documentclass[review]{elsarticle}

\usepackage{lineno,hyperref}
\usepackage{amsmath}
\usepackage{amssymb}
\usepackage{siunitx}
\usepackage{nth}
\usepackage{graphicx}
\usepackage{wasysym}
\usepackage{color} 
\modulolinenumbers[5]

\journal{Journal of \LaTeX\ Templates}
\usepackage{scalerel}
\usepackage{pbox}
\def\TransiTUM{\scalerel*{TransiTUM}{X}}
\def\MomenTUM{\scalerel*{MomenTUM}{X}}

\usepackage{setspace}

\begin{document}
\begin{frontmatter}

\title{A generic and density-sensitive method for multi-scale pedestrian dynamics}



\author[a]{Daniel H. Biedermann\corref{cor1}}
\ead{daniel.helmut.biedermann@gmail.com}
\author[a]{Jan Clever}
\author[a]{Andr\'{e} Borrmann}

\address[a]{Chair of Computational Modeling and Simulation, Technische Universit\"{a}t M\"{u}nchen, Arcisstrasse 21, M\"{u}nchen 80333, Germany}




\begin{abstract}
Microscopic approaches to the simulation of pedestrian dynamics rely on modelling the behaviour of individual agents and their mutual interactions. Regarding the spatial resolution, microscopic simulators are either based on continuous (SpaceCont) or discrete (SpaceDisc) approaches. To combine the advantages of both approaches, we propose to integrate SpaceCont and SpaceDisc into a hybrid simulation model. Such a hybrid approach allows simulating critical regions with a continuous spatial resolution and uncritical ones with discrete spatial resolution while enabling consistent information exchange between the two simulation models. We introduce a generic approach that provides consistent solutions for the challenges resulting from coupling diverging time steps and spatial resolutions. Furthermore, we present a dynamic and density-sensitive approach to detect dense areas during the simulation run. If a critical region is detected, the simulation model used in this area is dynamically switched to a space-continuous one. The correctness of the hybrid model is evaluated by comparison with a established simulator. Its superior computational efficiency is shown by runtime comparison with a standard microscopic simulation.on with the simulation results of other, well-established simulation models.
\end{abstract}

\begin{keyword}
pedestrian dynamics; transition; hybrid model; multi scale; zoom; generic modeling;
\end{keyword}
\cortext[cor1]{Corresponding author. Tel.: +49-89-289-25060; fax: +49-89-289-25051.}
\end{frontmatter}

\section{Introduction}
\subsection{The three spatial scales in pedestrian dynamics}
\label{intro1}
\begin{figure}
   	\centering
   	\includegraphics[width=1.0\linewidth]{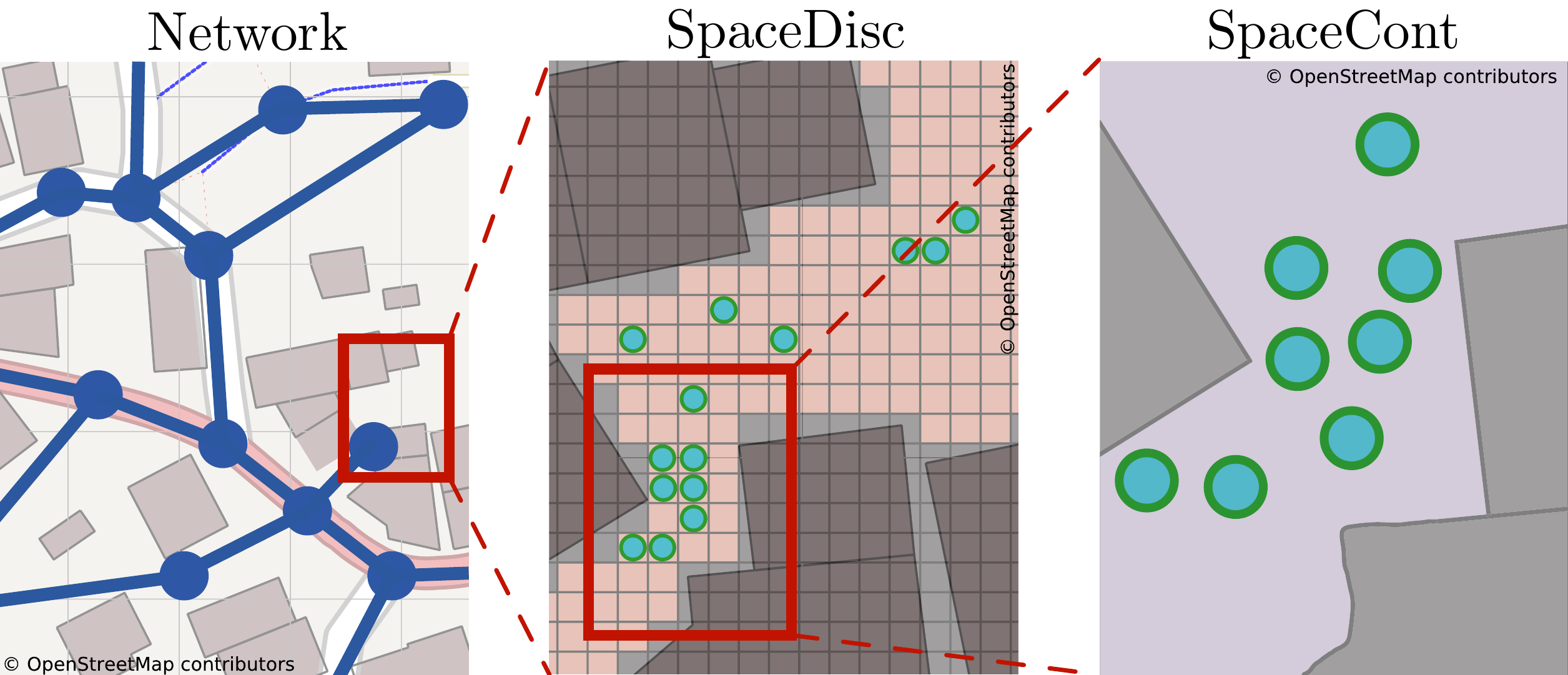}
   	\caption{The three spatial scales of pedestrian dynamics simulations: network models, spatial-discrete models and spatial-continuous models}
   	\label{fig:3Scales}
   \end{figure}

The simulation of pedestrian dynamics is required for a wide range of applications including the evacuation \cite{Ding201560, doi:10.1080/23249935.2018.1432717} or support of crowd management during public events \cite{Kluepfel2005}. A large number of different pedestrian dynamics models exist. We categorize these models into three different types according to their spatial resolution: spatial-continuous (SpaceCont) models, spatial-discrete (SpaceDisc) models and network models (see Figure \ref{fig:3Scales}).

Network approaches \cite{colombo2005pedestrian, hartmann2013structured} are often based on classic LWR-models \cite{lighthill1955kinematic,richards1956shock}. In these models, the simulation scenario is reduced to a network of nodes and edges. In consequence, these models have difficulties to simulate pedestrians in open spaces due to their one-dimensional nature. These models treat pedestrians not as individual subjects, but as cumulated pedestrian densities. Network models have low spatial resolution, but can be simulated with low computational costs. Therefore, these models are sufficient to simulate large, network-like scenarios (e.g. street networks).

SpaceDisc models simulate pedestrians as individual and discrete agents. SpaceDisc models are slower than network models, as they compute the behavior of individual subjects. The scenario is discretized by means of a two-dimensional grid. Consequently, the number of possible spatial positions a pedestrian can occupy is artificially restricted by the total number of grid cells. This discretization can lead to precision issues regarding the spatial resolution \cite{ZAWIDZKI201488}. These models are typically implemented by cellular automata \cite{Blue2001,Bandini2011}. Geometries of the grid's singular cells are mainly quadratic or hexagonal \cite{birch2007rectangular}. However, special forms such as triangular cells \cite{chen2014modeling} exist. Each cell can be occupied by a maximum of one person per cell \cite{schadschneider2009evacuation}. The spatial resolution is limited by the size of these grid cells.
 
 A spatial resolution, which is limited by the unit cell's size, is particularly problematic if the local pedestrian density gets too high. The maximal density, which can be simulated by SpaceDisc models is limited to $\rho = 1/A$, with $A$ as the area of a unit cell. We studied the usage of different spatial scales in previous projects \cite{Biedermann:2014:FBI, biedermann2014towards,kielar2015gentle, Biedermann:2016:MobilTUM,Biedermann:2016:FbiSkalen}. Table \ref{tab2} gives an overview about the most used geometric representations of pedestrians for spatial-discrete and spatial-continuous models as well as the maximum density that can be simulated.
	\begin{table}[h]
		\centering
		\caption{Geometries and maximal density values for spatial-discrete and spatial-continuous models from \cite{Biedermann:2016:FbiSkalen}. Size values of the torsos are based on Weidmann \cite{weidmann1992transporttechnik}. }
		{\tabcolsep14pt
			\begin{tabular}{c|c|c}
				\textbf{\it{SpaceDisc}}& \textbf{\begin{tabular}[c]{@{}c@{}}Minimal\\ Cell Size\end{tabular}} & \textbf{\begin{tabular}[c]{@{}c@{}}Maximal\\ Density\end{tabular}}  \\ \hline
				\textbf{\begin{tabular}[c]{@{}c@{}}Triangular\\ Cell Shape\end{tabular}} & $a_t=2\sqrt{3}r\approx\SI{0.80}{\meter}$                                                                  & $\frac{1}{A_t}=\frac{4\sqrt{3}}{3a^2_t}\approx\SI{3.64}{ped\per\meter\tothe{2}}$                                                                                                             \\ \hline
				\textbf{\begin{tabular}[c]{@{}c@{}}Quadratic\\ Cell Shape\end{tabular}}  & $a_q=2r=\SI{0.46}{\meter}$                                                                  & $\frac{1}{A_q}= \frac{1}{a^2_q}\approx\SI{4.73}{ped\per\meter\tothe{2}}$                                                                                                                       \\ \hline
				\textbf{\begin{tabular}[c]{@{}c@{}}Hexagonal\\ Cell Shape\end{tabular}}  & $a_h=\frac{2\sqrt{3}}{3}r\approx\SI{0.27}{\meter}$                                                                 & $\frac{1}{A_h}=\frac{2\sqrt{3}}{9a^2_h}\approx\SI{5.46}{ped\per\meter\tothe{2}}$                                                                 \\
				\hline\hline   
				\textbf{\it{SpaceCont}}& \textbf{\begin{tabular}[c]{@{}c@{}}Torso\\ Size\end{tabular}} & \textbf{\begin{tabular}[c]{@{}c@{}}Maximal\\ Density\end{tabular}}  \\ \hline      
				\textbf{\begin{tabular}[c]{@{}c@{}}Circular\\ Torso\end{tabular}} & $r=\SI{0.23}{\meter}$                                                                  & $\frac{1}{A_c}=\frac{\eta_c}{r^2\pi}\approx\SI{5.46}{ped\per\meter\tothe{2}}$                                                                                                                          \\ \hline    
				\textbf{\begin{tabular}[c]{@{}c@{}}Elliptical\\ Torso\end{tabular}} & \pbox{20cm}{$a=r$ \\ $b=\frac{1}{2}r$} \newline                                                                & $\frac{1}{A_e}=\frac{\eta_e}{a\cdot b\pi}\approx\SI{10.47}{ped\per\meter\tothe{2}}$                                                                           \\                                   
			\end{tabular}
		}
		\label{tab2}
	\end{table} 
   Higher densities cannot be represented by cell-based models due to their restricted spatial resolution. However, dense crowds are the greatest threat to the safety of pedestrians (every year, about 2000 people die due to dense human crowds \cite{hughes2003flow}). Consequently, SpaceDisc models are insufficient to simulate critical situations.
   
   Additionally, the calculation of velocities for SpaceDisc models is non-trivial due to their discretized nature. However, Bandini et al. made an approach to overcome this issue in modeling lower velocities by yielding movements for multiple time steps \cite{bandini2017approachw}.
   
\begin{figure}
    	\centering
    	\includegraphics[width=0.6\linewidth]{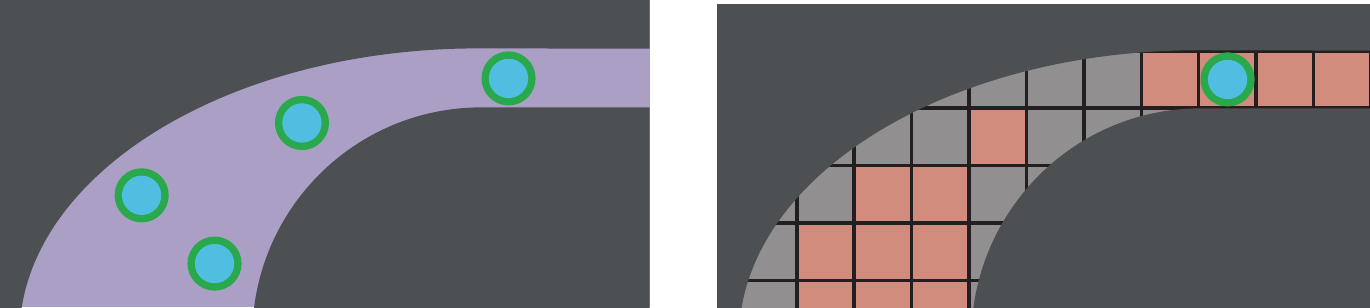}
    	\caption{On the continuous space, pedestrians pass a bottleneck (left side). The same narrow passage is not accessible for a pedestrian on the spatial discrete space (right side).}
    	\label{fig:bottleneck05}
    \end{figure}

   Another issue caused by SpaceDisc models is the simulation of areas with cells containing small-scale objects \cite{Biedermann:2016:FbiSkalen}. Such objects restrict the movement space of pedestrians in an unrealistic way, due to the limited spatial resolution of cellular automaton. This causes the problem that objects, which are smaller than the cell size, nevertheless occupy the total space of an entire cell. In the SpaceCont model (and in reality), a pedestrian can cross a bottleneck (e.g. a narrow passage) if the passage is wider than the pedestrian's torso. However, the low spatial resolution of SpaceDisc models can cause situations, in which a pedestrian cannot cross a bottleneck although the torso size is smaller than the bottleneck itself. The reason for this behavior is the artificial discretization with cells. If a cell contains an obstacle, the whole space of this cell is unaccessible for a pedestrian (see Figure \ref{fig:bottleneck05}).
   
    For these reasons, areas with high pedestrian densities or small scale obstacles have to be simulated by very precise simulation models to understand hazardous situations in detail. Pedestrian dynamics models with continuous and therefore unrestricted spatial resolution are most suitable for this task. Unfortunately, a high spatial resolution requires a high computational effort. Spatial-continuous models have higher computational costs than spatial-continuous models due to their infinite spatial state space. A spatial state space describes the total number of possible positions a pedestrian can take. In the case of a SpaceDisc model, a simulated agent can occupy any position on the two-dimensional plane, without artificial restrictions. By contrast, the spatial state space of spatial-discrete models is restricted by the total number of grid cells. Therefore, many calculations (e.g. spatial inquiries) are much slower for spatial-continuous simulation models. 
    
Well-known representatives of the spatial-continuous approach are the Social-Force models \cite{helbing1995social,helbing2000simulating, helbing2002simulation}. The movement of pedestrians in these models is based on the derived forces from a global potential field. Objects in the simulation scenario (e.g. pedestrians, walls or obstacles) have a repulsive potential, while the current target of a pedestrian has an attracting one. The total sum of the derived forces induces the movement of a pedestrian according to the classic equations of movement. These models should normally use state of the art wayfinding algorithms to have a plausible navigation behavior \cite{VIZZARI2020103241}

\subsection{Hybrid modeling in pedestrian dynamics} 
\label{backgroundInfoHybrid}
 In the context of this paper, we define hybrid modeling as the combination of pedestrian dynamics models from different spatial scales \cite{cristiani2014multiscale}, opposed to other hybrid approaches \cite{KNEIDL2013223}. In this manner, it is possible to simulate sub-sectors of the scenario by different spatial resolutions. In the proposed approach, special sections (e.g. areas with bottlenecks) are calculated by spatial-continuous models, while less significant parts (e.g. wide and open spaces) are simulated by spatial-discrete models. On that way, the computational effort compared to pure spatial-continuous simulations can be reduced \cite{chooramun2011implementing}. 
  
  A number of hybrid models have been proposed before in the field of pedestrian dynamics. Nguyen et al. coupled a LWR-model with an agent-based spatial-continuous model \cite{anh2012hybrid}. A hybrid model with three different spatial resolutions was developed by Chooramun et al. \cite{Chooramun2012}. Crociani et al. made a similar approach for an urban case study \cite{10.1007/978-3-319-51957-9_9}. Xiong et al. use transition cells between the coupled models to carry out the transition between different scales \cite{xiong2010hybrid}. Was and Lubas combine agent-based approaches with non-homogeneous and asynchronous cellular automata to simulate complex scenarios \cite{articlesdfgjtrt5}. The new concept of Crociani et al. is a hybrid approach which tries to replace the small scale model through an iterative search of optimal solutions \cite{inproceedingsCrociani}. Tordeux et al. describe pedestrian movement by means of aggregated density-flow relationships of individual pedestrian agents. This is results in a high efficient and high accurate hybrid model \cite{TORDEUX2018128}. An overview about current approaches for hybrid modeling in the field of pedestrian dynamics can be found by Ijaz et al. \cite{Ijaz2015}.
  
   Unfortunately, many hybrid models share one common issue. They are tailored for the coupling of very specific individual models and lack genericity. So far, generic hybrid approaches are still missing in the field of pedestrian dynamics \cite{Ijaz2015}. Researchers willing to combine two models from different scales must develop a new hybrid coupling scheme for each specific pair of models. For example, a research team has to develop one hybrid model for coupling the spatial-continuous model by Helbing and Molnar \cite{helbing1995social} with the spatial-discrete model by Blue and Adler \cite{Blue2001}. If the spatial-discrete model is later replaced by an alternative cellular automaton (e.g. \cite{Bandini2011}), a new coupling scheme has to be developed, since different models have varying requirements for a successful coupling. 
  
  However, if we understand the fundamental and universal concepts of transformation conditions, we can develop a general approach, which allows to combine  pedestrian dynamics models from different scales. The \TransiTUM{} approach follows this idea and generalizes the necessary conditions for the transformation of models between different scales. Consequently, coupling of arbitrary pedestrian dynamics models is possible with the method presented here \cite{biedermann2014towards, Ijaz2015}. 
  
  In currently available hybrid models, some parts of a scenario can be simulated by a more detailed model, while other parts are simulated in less detail. However, if one part of a scenario is simulated by a specific model it will be computed by this model during the whole simulation run. This leads to a number of significant issues, since the main scope of the scenario can be time dependent. An example for this case is the simulation of a music concert.  At the beginning of such an event, our main scope should be the entry, later on the areas around the stages have a higher relevance. However, at the end of the event, when people leave the concert, our scope should be on the exit areas. Without a possibility to switch dynamically and locally between the used models, all the mentioned locations have to be simulated with a high resolution model. However, with a dynamic zoom approach, we are able to assign dynamically which areas of a scenario have to be simulated in detail and which not. This leads to more precise results.
  
  The selection of scope works automatically and is based on the local density. Doing so, we can ensure that all potential dangerous situations are calculated with a detailed enough model. Such a dynamic zoom approach for hybrid modeling is currently unknown in the field of pedestrian dynamics. To overcome this issue, we extended the generic \TransiTUM{} approach by such a dynamic zoom concept.

\section{Generic transition approach}
\label{sec:genTransApproch}

 \subsection{Overview about the methodology}
\begin{figure}
	\centering
	\includegraphics[width=1\linewidth]{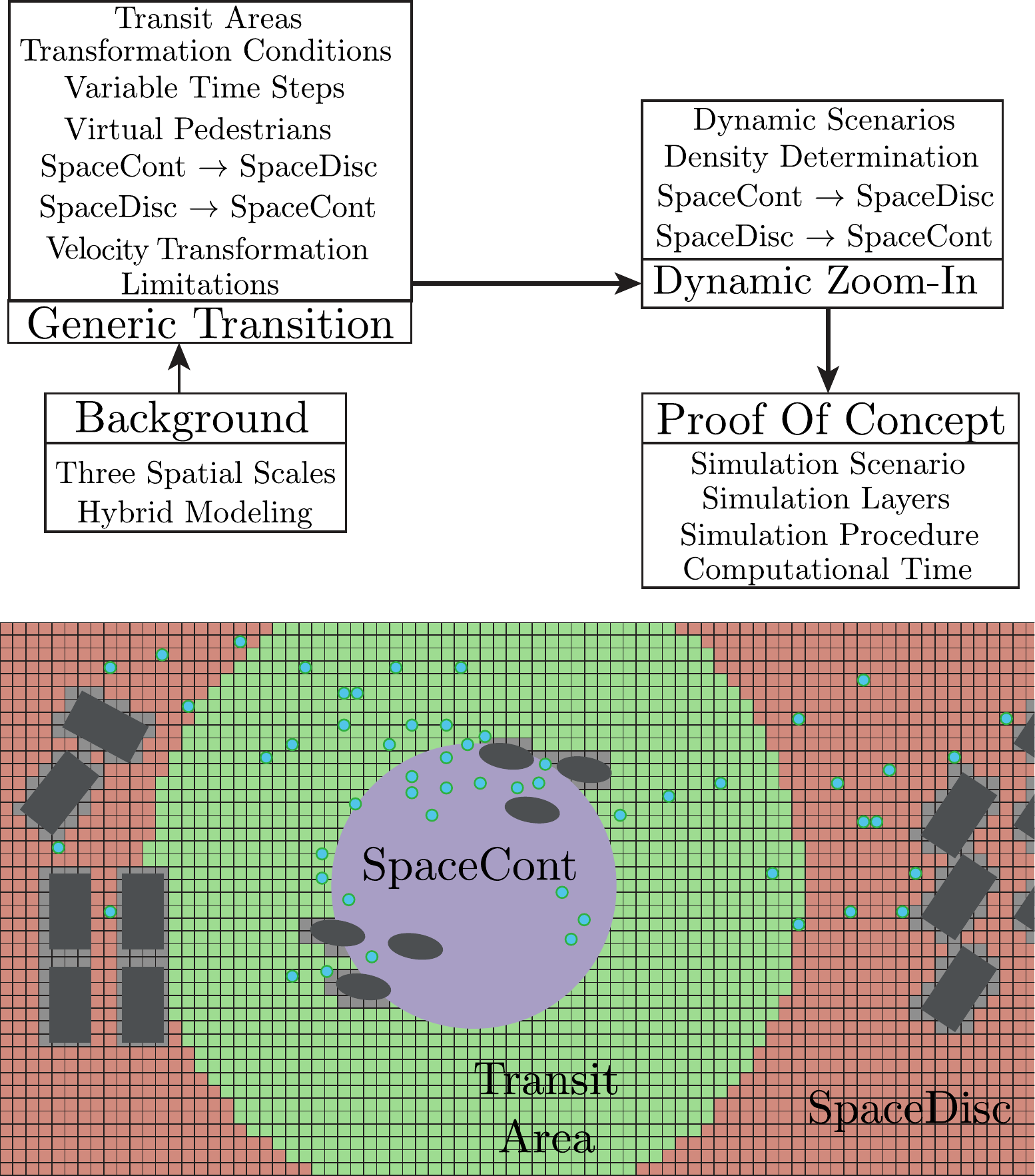}
	\caption{Overview of the methodology and the different simulation areas. Multiscale simulation scenario with the spatial-continuous area (violet color), spatial-discrete area (orange color) and the transit area (green). Exemplary pedestrians are shown as blue circles with green, circular edges.}
	\label{fig:overview03}
\end{figure}

The \TransiTUM{} methodology aims at fulfilling a generic transition approach combined with a dynamic and density depending zoom-in. A overview of our methodology is shown in Figure \ref{fig:overview03}.

At first, we provided background information about the three spatial scales (see Section \ref{intro1}) as well as hybrid modeling in the context of pedestrian dynamics (see Section \ref{backgroundInfoHybrid}).

In the following Chapter \ref{sec:genTransApproch} we explain the procedure to transform pedestrians from a spatial-discrete model to a spatial-continuous one and vice versa. Therefore, we describe the principle structure of transit areas (see Section \ref{subsec:TheTransiZOnes}) and the necessary boundary conditions for our generic approach (see Section \ref{seubsec:NecCondForThGenTransAP}). Different simulation models can have various simulation time step durations. How 
we treat this issue is described in Section \ref{subsec:Time}. Moreover, the interaction between pedestrians of different scales in the shared transit areas is handled by a virtual pedestrian approach (see Section \ref{sec:virtualPeds}). In Sections \ref{sec:TranMi2Me} and \ref{sec:TraMe2Mi} we describe the actual transformation procedure in these zones. At the end of this chapter, we describe the limitations of our approach (see Section \ref{sec:Limit}).

The automatic and dynamic Zoom-In approach for high density situations is described in Chapter \ref{sec:DynamicZoom}. In Section \ref{deTrSec:31} we explain the motivation for a dynamic transformation approach, while Section \ref{deIFF:32} describes a method to determine pedestrian densities in an automatic way. Sections \ref{sec:ZoomIn} and \ref{sec:ZoomOut} define the actual Zoom-In and Zoom-out approach.

Chapter \ref{cdtfasd4} gives a proof of concept for the functionality of \TransiTUM{}. In Section \ref{cdtfasd4o1} we describe the event scenario for our simulation. The used simulation layers are explained in Section \ref{cdtfasd4o2} and the procedure is executed in the following Section \ref{cdtfasd4o4}. The amount of saved computational effort is given in Section \ref{cdtfasd4o4}.
The last Chapter \ref{OutlookandConclusionas} gives a summary as well as an outlook to future work.

\subsection{The transit areas}
\label{subsec:TheTransiZOnes}
The \TransiTUM{} method allows the coupling of spatial-continuous and spatial-discrete simulation models by implementing a spatial decomposition approach. This means, that dedicated areas of the simulation are either simulated by spatial-continuous or spatial-discrete approaches and the simulated pedestrians are handed over from one simulation model to the other. To allow this hand-over, a transition both in time and in space is required. The transition is realized within a transit area in which both simulation models are active. Pedestrian agents that leave the zone associated to a certain simulation model are removed from it, while agents entering a zone are newly generated. Only in the transit area, the agents are subject to both coupled simulation models.

\TransiTUM{}'s hybrid scenario structure is depicted in Figure \ref{fig:overview03}. Each simulation model has full knowledge about all static objects of the entire scenario (e.g. all obstacles and walls), but only knowledge about those dynamic objects (e.g. pedestrians) which are in the area associated with their simulation model. The transformation of a pedestrian between different spatial scales is carried out by \TransiTUM{} in the transit area. All SpaceCont areas are surrounded by such zones. 

Consequently, if a pedestrian agent leaves the area of its current simulation model, it will enter a transit area. These zones ensure that all pedestrians who leave the region of their current simulation model are correctly transformed from spatial-continuous to spatial-discrete or vice versa. Each transit area has knowledge about all static and all dynamic simulation objects, which are placed inside its area. Therefore, spatial-continuous and spatial-discrete pedestrians are simulated simultaneously in the transit areas. 

 Pedestrians can be transformed only if they are in the region of the transit area. Consequently, this zone has to be wide enough that no dynamic simulation object can cross the transit area without being transformed. The minimal necessary width can be determined as follows: Pedestrian dynamics simulations are performed in discrete simulation time steps. For each time step, the simulation calculates and updates the new positions of all pedestrians. The shorter the time step of a simulation, the higher is its resolution in time. We have two simulation models and therefore two different time step durations: A spatial-discrete time step $\Delta t_{disc}$ and a spatial-continuous  time step $\Delta t_{cont}$. A higher spatial resolution requires a higher resolution of time (see Section \ref{subsec:Time}):
\begin{equation}
\Delta t_{disc} \geq \Delta t_{cont}
\end{equation}
As stated above, we have to ensure that pedestrians can not move directly from the region of the current simulation model to a region of another simulation model during one simulation time step. Otherwise, the \TransiTUM{} model is not able to execute the transformation process for these simulated pedestrian agents. This means, the transit areas have to be wider than the distance any pedestrian could walk during one simulation time step:
\begin{equation}
w_{Tr} > v_{max}\cdot\Delta t_{disc}
\end{equation} 
According to Weidmann \cite{weidmann1992transporttechnik}, the maximal velocity of a pedestrian is given by $v_{max}=\SI{2.16}{\metre\per\second}$. The positions of the individual transit areas can change during the simulation run by the use of a dynamic zoom approach (see Chapter \ref{sec:DynamicZoom}).

\subsection{Necessary conditions for the generic transformation approach}
\label{seubsec:NecCondForThGenTransAP}
Some general conditions have to be fulfilled by the spatial-continuous and spatial-discrete models to realize a generic transition by the \TransiTUM{} framework. Spatial-continuous models are simulated on a continuous space and spatial-discrete models on a discretized cellular grid. A specific set of simulation objects is necessary to enable a successful generic combination of these models (see Table \ref{rat}). The \TransiTUM{} framework needs to know and influence these variables and parameters for the transition procedure. 

Each individual model needs an individual and discrete representation of a pedestrian $p_i, p_j$. Both models need global identification numbers $i,j$ for their pedestrian agents $p_i,p_j$. Furthermore, both models need a constant duration of simulation time steps $\Delta t_{cont},\Delta t_{disc}$ during the whole simulation run. Additionally, a simulated pedestrian agent $p_i,p_j$ has a current velocity $\vec{v}_i,\vec{v}_j$. These simulation elements have to be available in both models. Since the spatial-discrete model simulates on a cellular grid, cells $\bf{c_{m,n}}$ with a cell center position at $\vec{c}_{m,n}$ are mandatory. The indexes $m$ and $n$ describe the row and column numbers of the cell $\bf{c_{m,n}}$ from cellular grid. 

In the case of cells with quadratic geometries, the numbering of rows and columns is trivial. However, in the case of hexagonal shaped cell geometries more complex numbering methods can be necessary \cite{HexGridsAmPa}. The position $\vec{p}_j$ of a spatial-discrete pedestrian $p_j$ is given by the cell center position $\vec{c}_{m,n}$ of her current cell $\bf{c_{m,n}}$. For the spatial-continuous model, a radius of the pedestrian's torso $r_i$ is necessary. In the case of elliptical formed pedestrian agent geometries, the radius of the semi-major axis is used as the torso radius. The position of a spatial-continuous pedestrian is described by the vector $\vec{p}_i$. 
\begin{table}[htbp]
\centering
		\caption{Overview of mandatory simulation elements for the generic coupling of spatial-continuous and spatial-discrete models}

		\begin{tabular}{|p{60pt}|p{180pt}|p{66pt}|}
			\hline \bf{Elements} & \bf{Explanation} & \bf{Model Type}\\
			\hline $p_i, p_j$& individual and discrete pedestrian & both\\
			\hline $i, j$&indices for a spatial-continuous $i$ and a spatial-discrete $j$ pedestrian& both\\
			\hline $\Delta t_{cont},\Delta t_{disc}$& constant duration of simulation time steps & both\\
			\hline $\vec{v}_{i}, \vec{v}_{j}$& current velocity of pedestrian $p_i, p_j$ & both\\
			\hline $\bf{c_{m,n}}$ & cells $\bf{c_{m,n}}$ of a grid $\bf{C}$, which are empty, occupied by obstacles or by a pedestrian & spatial-discrete\\
			\hline $\vec{c}_{m,n}$& cell center position of a cell $\bf{c_{m,n}}$& spatial-discrete\\
			\hline $A_{m,n}$ & Area of a cell  $\bf{c_{m,n}}$ & spatial-discrete\\
			\hline $\vec{p}_{j}$& cell center position of the current cell from pedestrian $p_j$ & spatial-discrete\\
			
			\hline $r_i$ & torso radius of pedestrian $p_i$ & spatial-continuous\\
			\hline $\vec{p}_i$ & current position of pedestrian $p_i$ & spatial-continuous\\			
	
			\hline
		\end{tabular}
		\label{rat}	
\end{table}

In addition to the presence of these simulation elements, the following axioms must hold to enable the generic transition approach:
\begin{enumerate}
	\item An upper boundary $v_{max}$ exists for the velocities $v_i,v_j$ of both simulation models.
	\item All grid cells $\bf{c_{m,n}} \in \bf{C}$ are convex and have the same geometric shape and size. This results in a regular and homogeneous grid.
	\item A cell $\bf{c_{m,n}}$ can be occupied by a maximum of one pedestrian. Therefore, the grid cell area $A_{m,n}$ has to be large enough to contain a spatial-continuous pedestrian.
	\item The constant time step of the spatial-continuous model is smaller than the spatial-discrete simulation time step: $\Delta t_{cont}\leq\Delta t_{disc}$ (see Table \ref{tab:MiMeTiSt})
\end{enumerate}
The described simulation elements and necessary axioms are normally given in any SpaceCont or SpaceDisc model. Consequently, \TransiTUM{} can be used as a generic transition framework to couple arbitrary pedestrian dynamics models.

\subsection{Variable time step duration}
\label{subsec:Time}
Pedestrian dynamics models are mainly simulated in discrete and constant simulation time steps. After each simulation time step, the positions of all pedestrians are recalculated based on the underlying simulation model. Consequently, the simulation time step defines the resolution of time. For example, if a scenario, which lasts two hours, is simulated by constant simulation time steps of one second, the simulator would have to determine the positions of all pedestrians 7200 times. If we lower the time resolution to a simulation time step of two seconds, only 3600 calculation runs are necessary. So for the cost of time resolution, we increase the computational efficiency of our simulator. Therefore, it is important to find the ideal middle course between computational efficiency and accuracy of the computed results.

The duration of one time step depends on the spatial resolution of the used simulation model. A good example of this issue is described by Westervelt \cite{westervelt1998simulating}. The spatial resolution of a simulation model is qualified by the minimum recognizable length $\Delta x_{res}$.

 For example, a cellular automaton based on quadratic cells with edge size $a$ would have a recognizable length of $\Delta x_{res} = a$. This means, that the best possible spatial resolution is determined by the size of the cells:
\begin{equation}
\Delta t = \frac{\Delta x_{res}}{v_{max}}
\end{equation}
This equation is well-known as the Courant-Friedrichs-Lewy condition \cite{Courant1928}. For pedestrian dynamics simulation with SpaceDisc models, the size of a cell has to be large enough to contain one pedestrian (see Table \ref{tab2}). This represents the spatial resolution $\Delta x_{res}$. The maximal velocity of pedestrians depends on the simulated scenario (see Table \ref{tab:herdingFactor}). Additionally, this equation can be used for SpaceCont models. In this case, the spatial resolution $\Delta x_{res}$ is given by the size of the smallest, not ignorable object in the scenario. Ignorable means, that a object can not be passed by a pedestrian.
\begin{table}
	\centering
	\caption{Scenario depending Velocities according to Weidmann \cite{weidmann1992transporttechnik}}

	\begin{tabular}{c c}
		\hline\noalign{\smallskip}
		Scenario             & \multicolumn{1}{c}{Average} \\
		& \multicolumn{1}{c}{Velocity \cite{weidmann1992transporttechnik}}\\\hline	
		Commercial Traffic & $\SI{1.45}{\metre\per\second}$ to $\SI{1.61}{\metre\per\second}$\\
		Commuter Traffic  &$\SI{1.34}{\metre\per\second}$ to $\SI{1.49}{\metre\per\second}$       \\		
		Shopping Traffic &$\SI{1.04}{\metre\per\second}$ to $\SI{1.16}{\metre\per\second}$\\
		Public Event  &$\SI{0.99}{\metre\per\second}$ to $\SI{1.10}{\metre\per\second}$ \\			
		
	\end{tabular}
	\label{tab:herdingFactor}       
\end{table}

In theory, a spatial-continuous model has an indefinitely small spatial resolution and therefore indefinitely short time steps. In real simulators, time steps of spatial-continuous models are discretized and their durations have lower limits.

Unfortunately, the current literature does not suggest a minimal necessary duration for time steps of spatial-continuous simulation models. Therefore, we performed a literature review to find a minimal necessary time step duration. We studied various pedestrian dynamics models to find this lower boundary. The results are shown in Table \ref{tab:MiMeTiSt}. The lowest boundary we found were spatial-continuous pedestrian dynamics models with a time step duration of $\SI{0.01}{\second}$ \cite{helbing2000simulating,HlebingWeb,PhysRevE.82.046111}. 

In conclusion, a minimal time step duration of $\SI{0.01}{\second}$ seems to be sufficient to simulate any pedestrian dynamics model. Furthermore, our study revealed that the time step duration of spatial-discrete models is larger than the time step duration of any spatial-continuous approach.
\begin{table}[htbp]
	\centering
	\caption{Overview of different pedestrian dynamics approaches and their used simulation time step duration}
	
	\begin{tabular}{|p{35pt}|p{125pt}|p{60pt}|}
		\hline \multicolumn{3}{|c|}{\bf{Spatial-Continuous Models}} \\
		\hline \bf{Work} & \bf{Authors}  & \bf{Time step}\\
		\hline \cite{PhysRevE.82.046111} & Chraibi et al. & \SI{0.01}{\second}\\
		\hline \cite{helbing2000simulating,HlebingWeb} & Helbing et al. & \SI{0.01}{\second}\\
		\hline \cite{Kwak2016} &Kwak et al. & \SI{0.05}{\second}\\
		\hline \cite{moussaid2010walking} &Moussa{\"i}d et al. & \SI{0.05}{\second}\\		
		\hline \cite{alighadr2013simulation} & Alighadr et al.  & \SI{0.10}{\second}\\
		\hline \cite{Dai20132202} & Dai et al. &\SI{0.10}{\second}\\
		\hline \cite{okazaki1993study} &Okazaki and Matsushita  & \SI{0.10}{\second}\\
		\hline \cite{PhysRevE.72.026112} & Yu et al.c & \SI{0.05}{\second} - \SI{0.15}{\second}\\
			\hline \multicolumn{3}{|c|}{\bf{Spatial-Discrete Models}} \\
			\hline \bf{Work} & \bf{Authors} & \bf{Time step}\\		
		\hline \cite{Bandini2011} & Bandini et al.&  \SI{0.30}{\second}\\
		\hline \cite{Kirchner2002260} &Kirchner and Schadschneider&  \SI{0.30}{\second}\\
		\hline \cite{Bandini2006} & Bandini et al. & \SI{0.33}{\second}\\
		\hline \cite{algadhi1991simulation} &  AlGadhi and Mahmassani &  \SI{0.36}{\second}\\
		\hline \cite{Chen2015287} & Chen et al.& \SI{0.40}{\second}\\
		\hline \cite{Weifeng2003633} & Weifeng et al.&  \SI{0.40}{\second}\\
		\hline \cite{Xiong2013} & Xiong et al.&  \SI{0.50}{\second}\\
		\hline \cite{Yamamoto2007654} & Yamamoto et al.&  \SI{0.50}{\second}\\
		\hline \cite{Antonini2006667} & Antonini et al.& \SI{0.90}{\second}\\
		\hline \cite{Blue2001} & Blue and Adler& \SI{1.00}{\second}\\

		\hline
	\end{tabular}
	\label{tab:MiMeTiSt}	
\end{table}

Based on our investigations, we conclude that simulators on heterogeneous scales have diverging time step durations. This issue has to be considered by a generic transformation approach. A simulation is executed over a simulated time interval of $t\in\left[0,t_{end}\right]$. At the beginning of a hybrid simulation, at $t=0$, the spatial-continuous as well as the spatial-discrete model are in their start configuration. This means, that the pedestrians are handled by both models at exactly the same point in time at $t=0$. However, the following simulation time steps of these models are simulated at different points in time as the models may have diverging time step durations. 

Unfortunately, a timely-coherent transformation of models is possible only if the pedestrians' positions are calculated at the same point in time. Nevertheless, the transformation process can be executed easily, if the spatial-discrete time step $\Delta t_{disc}$ is an integral multiple of the spatial-continuous time step $\Delta t_{cont}$:
\begin{equation}
\exists k\in\mathbb{N}:  k\cdot \Delta t_{cont} = \Delta t_{disc}
\label{eq:inmathbb}
\end{equation}

In case Equation \ref{eq:inmathbb} holds, the transformation can be directly performed after the calculation of a spatial-discrete $\Delta t_{disc}$ and its corresponding spatial-continuous time steps $k\cdot\Delta t_{cont}$. 

For example, the spatial-discrete time step $\Delta t_{disc} =  \SI{1.00}{\second}$ of the Blue and Adler model \cite{Blue2001} is an integral multiple of the  spatial-continuous time step $\Delta t_{cont} = \SI{0.05}{\second}$ duration of the Moussa{\"i}d et al. model \cite{moussaid2010walking}. Therefore, a transformation can be easily executed after the calculation of one spatial-discrete time step respectively twenty spatial-continuous time steps. However, an extrapolation is necessary if the singular time step durations are not an integral multiplicity of each other:
\begin{equation}
\nexists k\in\mathbb{N}:  k\cdot \Delta t_{cont} = \Delta t_{disc}
\label{eq:inmathbb2}
\end{equation}

This results in a time gap between the spatial-discrete time step duration $\Delta t_{disc}$ and the corresponding spatial-continuous time steps $k\cdot\Delta t_{cont}$:
\begin{equation}
0<\Delta t_{disc}-k\cdot\Delta t_{cont} < \Delta t_{cont}
\label{timediscrepency}
\end{equation}

\begin{figure}
	\centering
	\includegraphics[width=0.8\linewidth]{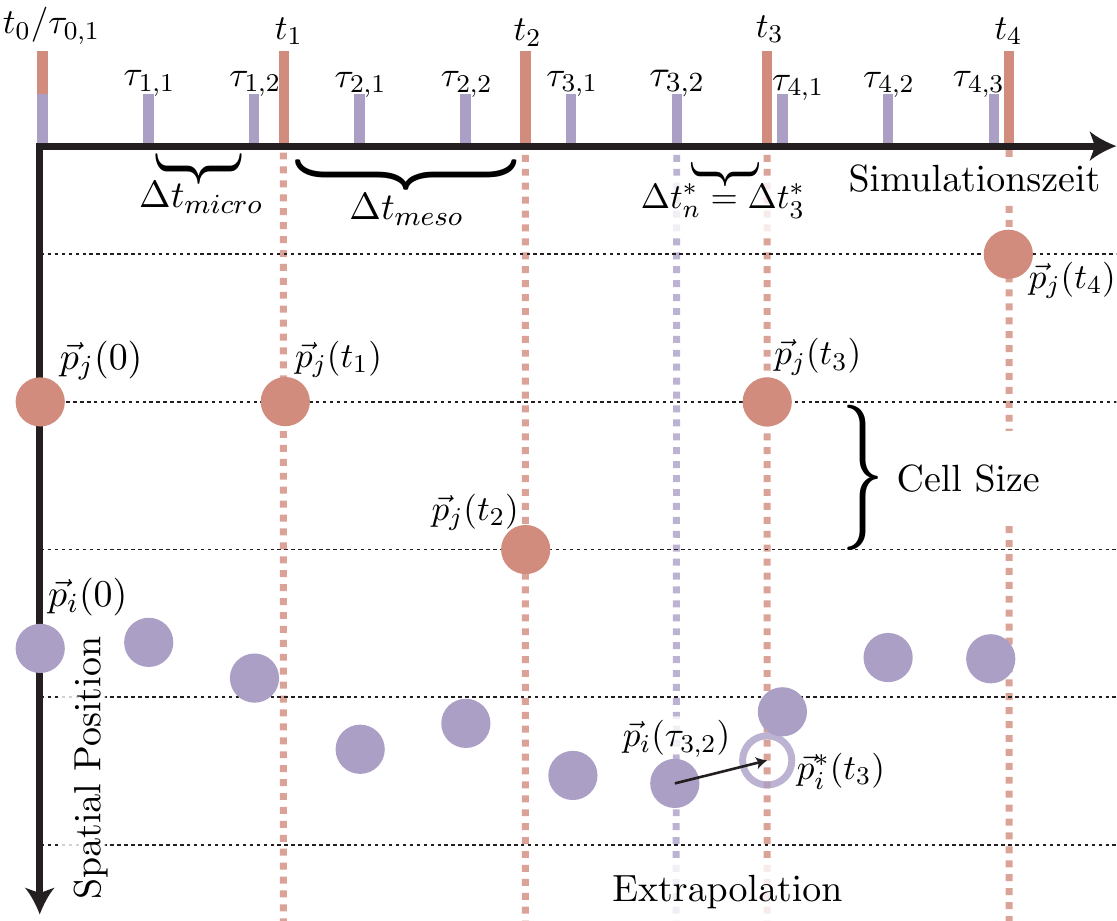}
	\caption{Time-space-diagram for a spatial-discrete pedestrian $p_j$ and a spatial-continuous pedestrian $p_i$. The time axis is discretized by the spatial-discrete time steps $t_n$ and the spatial-continuous time steps $\tau_{n,l}$. Exemplary, the extrapolation is shown for $n=3$.}
	\label{fig:SpaceTimeGraph_10}
\end{figure}

After the execution of a number of time steps, the time differences add up to a complete spatial-continuous time step. Therefore, an additional spatial-continuous time step has to be executed between two spatial-discrete time steps. This variable number of $d_n$ spatial-continuous simulation time steps between two spatial-discrete time steps $\Delta t_{disc}$ can be determined by the following equation:
\begin{equation} 
d_n=\left\lfloor n\frac{\Delta t_{disc}}{\Delta t_{cont}}\right\rfloor-\left\lfloor (n-1)\frac{\Delta t_{disc}}{\Delta t_{cont}}\right\rfloor
\label{eq:dnTimesteps}
\end{equation}
The running index $n$ describes the consecutive numbering of the current time step $\Delta t_{disc}$. Between two transformation processes, the $n^{\text{\tiny th}}$ spatial-discrete time step and the corresponding $d_n$ spatial-continuous time steps have to be simulated. 

Referring to Equation \ref{timediscrepency}, the timing discrepancy between spatial-discrete and spatial-continuous models for the transformation process is always smaller than $\Delta t_{SpaceCont}$. This difference can be further minimized by the use of a linear extrapolation method. For the transformation calculations only, the current positions $\vec{p}_i$ of all spatial-continuous pedestrians $p_i$ in the transit areas are extrapolated according to their current velocity $\vec{v}_i$: 
\begin{equation}
\vec{p}_i^*(t_n) = \vec{p}_i(\tau_{n,d_n}) + \vec{v}_i\cdot\underbrace{\left( n\cdot\Delta t_{disc} -\left\lfloor n\frac{\Delta t_{disc}}{\Delta t_{cont}}\right\rfloor \cdot\Delta t_{cont}\right)}_{\Delta t^*_{n}}
\end{equation}

This extrapolation maps all spatial-continuous pedestrians in the transit areas on the point in time of the next spatial-discrete simulation time step. Consequently, the transformation process can be executed at this moment in time. An exemplary time-space-diagram is shown in Figure \ref{fig:SpaceTimeGraph_10}.

\subsection{Virtual pedestrians}
\label{sec:virtualPeds}
Virtual pedestrian are a necessary precondition to full-fill a generic transformation approach (see Figure \ref{fig:overview03})

If the \TransiTUM{} model is used, no interaction between the combined individual models (SpaceDisc and SpaceCont) is necessary for the transformation process. The complete communication process is executed solely by the \TransiTUM{} framework. Accordingly, the simulation of the spatial-discrete and the spatial-continuous model in the transit area is challenging. Both models are simulated simultaneously on the same area -- the transit area. As the coupled models do not interact with each other, the pedestrians of the respective other model are not considered. 

This means, that a spatial-discrete pedestrian has no knowledge about the existence of any spatial-continuous pedestrians and vice versa. Therefore, it is possible that the positions of two pedestrians from different scales overlap each other, i.e. they collide. Such unrealistic behaviour in the transit area can be avoided by the use of virtual pedestrians. A similar approach was used by L\"ammel et al. \cite{lammel2014large}.

Virtual pedestrians effect the simulation the same way normal simulated pedestrians influence each other. For example, a virtual pedestrian in a Social-Force model has the same repulsive forces like normal simulated pedestrian. However, virtual pedestrians do not move during the time they exist in the simulation. Their only purpose is to prevent overlaps of pedestrians in the transit area. Table \ref{tabVirutalPedestrians} shows an overview of the differences between normal simulated and virtual pedestrians.
	\begin{table}[h]
		\centering
		\caption{Differences between normal simulated and virtual 
        pedestrians}
		{\tabcolsep14pt
			\begin{tabular}{c|c|c}
				\textbf{\it{}}& \textbf{\begin{tabular}[c]{@{}c@{}}Normal\\ Pedestrian\end{tabular}} & \textbf{\begin{tabular}[c]{@{}c@{}}Virtual\\ Pedestrian\end{tabular}}  \\ \hline
				\textbf{\begin{tabular}[c]{@{}c@{}}Control\end{tabular}} & simulation model                                                                  & \TransiTUM{} model                                                                                                             \\ \hline
				\textbf{\begin{tabular}[c]{@{}c@{}}Effective Area\end{tabular}}  & {\begin{tabular}[c]{@{}c@{}}model area \& \\ transit area\end{tabular}}                                                                 & transit area                                                                                                                     \\ \hline
				                                              \textbf{\begin{tabular}[c]{@{}c@{}}Locomotion\end{tabular}}  & dynamic object & static object
			\end{tabular}
		}
		\label{tabVirutalPedestrians}
	\end{table} 

 According to the execution order for a given time frame, the positions of the pedestrians in the SpaceDisc model are calculated before they are simulated with the SpaceCont model. One objective of introducing virtual pedestrians is to avoid the collision of pedestrians during the simulation of the SpaceDisc model in the transit areas. The positions of all spatial-continuous pedestrians in the transit areas during the last executed spatial-continuous time step have to be passed to the SpaceDisc model. All cells, which contain a spatial-continuous pedestrian at the  $d_n^{\text{\tiny th}}$ time step are occupied by virtual pedestrian and thus closed for the spatial-discrete model. The cells are accessible again, after the execution of the $t_{n}^{\text{\tiny th}}$ time step.
 
 The inclusion of virtual pedestrians significantly increases the accuracy and precision of a multi-scale simulation approach. However, for the use of virtual pedestrians some additional boundary conditions have to be fulfilled for the presented generic hybrid simulation. In particular, the following conditions have to be met:
 
 \begin{enumerate}
 	\item The combined models have to allow the \TransiTUM{} framework to create virtual pedestrians, which do not move and are handled like static objects with the same characteristics as a normal simulated pedestrian. 
 	\item The calculation of the SpaceDisc time step $t_n$ has to be executed before the corresponding $d_n$ time steps $\tau_{n,1} ... \tau_{n,d_n}$ from the SpaceCont model are simulated.	
 \end{enumerate}
Since these are pure technical conditions, nearly any pedestrian dynamics model can be expanded to fulfill both. If these conditions are provided, a hybrid simulation with virtual pedestrians can be executed.

\subsection{Transformation from the SpaceCont model to the SpaceDisc model}
\label{sec:TranMi2Me}
In this section, we describe the transformation of a pedestrian $p_i$. The bi-directional transformation is the elemental part of our generic transformation approach.
The transformation process itself can be executed only in the transit areas (see Figure \ref{fig:overview03}). The transformation process takes place after a simulation time step $\Delta t_{disc}$ is calculated. We determine which pedestrians in the transit areas have to be transformed for the next simulation time steps. This is done by determining the propagation vector for all pedestrians located in the transit areas: 
\begin{equation}
\vec{\sigma}_i = \vec{e}_{v_i}\cdot v_{max}\cdot \Delta t_{disc}
\end{equation}
The unit vector $\vec{e}_{v_i}=\frac{\vec{v}_i}{v_i}$ describes the current walking direction of a pedestrian $p_i$. The total term describes the maximal distance a spatial-continuous pedestrian $p_i$ is able to walk during the time frame of the next time step $\Delta t_{disc}$ if the pedestrian keeps her current walking direction. We obtain the maximal reachable position $\vec{p}^*_i$  if we add the propagation vector to the current position $\vec{p}_i$ of this pedestrian.
\begin{equation}
\vec{p}^*_i = \vec{p}_i +  \vec{\sigma}_i
\end{equation}

The new position $\vec{p}^*_i$ respectively the cell containing this position is the desired position of the transformed pedestrian.

A pedestrian $p_i$ has to be transformed if the propagated position $\vec{p}^*_i $ is located outside of the transit area in the region of the SpaceDisc model. The velocity is $v_i$ of the transformed pedestrian equals the former velocity $v_j$. Due to the convex shape of the transit areas, it is guaranteed that a pedestrian, who keeps her walking direction can not leave the transit area during the next time step $\Delta t_{disc}$. Unfortunately, this approach is problematic if a pedestrian changes her current walking direction $\vec{e}_{v_i}$ during this time frame. In this case, it is possible that the pedestrian leaves the transit area although the propagated position is still in the region of her current transit area. For this reason, it is mandatory to consider direction changes in our approach \cite{Biedermann:2016:PED16}. According to Antonini et al. \cite{Antonini2006667}, pedestrians do not change their direction immediately, but stride by stride. They observed, that the average angle change per stride equals $\omega_{s}=\pm\ang{12.3}$. The human stride length depends on the pedestrian's current velocity and varies between $\SI{0.5}{\meter}$ for small and $\SI{2.2}{\meter}$ for large velocities \cite{weidmann1992transporttechnik}. Referring to Margaria \cite{margaria1976biomechanics}, Weidmann et al. \cite{weidmann1992transporttechnik} describe the relationship between the pedestrian's current velocity $v_i$ and the corresponding stride length $d_s$:
\begin{equation}
d_s(v_i) = \SI{0.234}{\meter}+\SI{0.302}{\meter}\cdot v_i
\end{equation}
We can calculate the maximal possible angular change $\pm\Omega_i$ a pedestrian can make during the next time step $\Delta t_{disc}$. This value depends on the maximal number of strides during this time frame:
\begin{equation}
\pm\Omega_i = \pm\frac{\Delta t_{disc}\cdot v_i}{d_s(v_i)}\cdot |\omega_s|
\end{equation}
\begin{figure}
\centering
\includegraphics[width=0.6\linewidth]{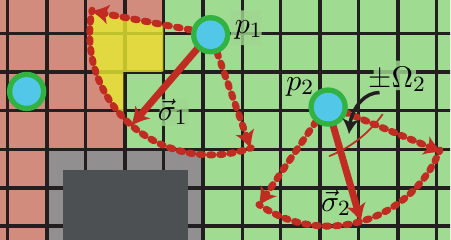}
\caption{Propagation vectors $\vec{\sigma}_i$ and angle $\pm\Omega_i$ to describe the circular propagation segments of pedestrian $p_i$. A pedestrian gets transformed if her propagation segment crosses the region of the SpaceDisc model (yellow marked area).}
\label{fig:MicroMesoPropagation_05}
\end{figure}
The maximal angular change is limited by $\pm\Omega_{max} =\pm\pi$, since the whole angular space is covered by the interval $[0,2\pi]$. A maximal angular change $\Omega_{max}$ means that a pedestrian has done a total u-turn compared to her previous walking direction $\vec{e}_{v_i}$. The angle $\pm\Omega_i$ describes a circle segment with radius $|\vec{\sigma}_i|$ around the current position $\vec{p}_i$ of the pedestrian. A pedestrian needs to be transformed if any part of this circle segment is located in the area of the SpaceDisc model. Figure \ref{fig:MicroMesoPropagation_05} shows an example of such circular propagation segments. Pedestrian $p_1$ has to be transformed, since a part of her propagation segment overlaps with the SpaceDisc area. Contrary, pedestrian $p_2$ will not be transformed, since she will stay in the transit area during the next time step $\Delta t_{disc}$.

For the transformation process, only these positions should be considered, which are reachable by the pedestrian. Therefore, we use the underlying grid from the SpaceDisc model to calculate all cells, which are reachable by the pedestrian. This can be achieved in principle by flooding algorithms (see \cite{hartmann2010adaptive}).These calculations can be quite time consuming for larger scenarios. Therefore, only cells within the propagation segments of the pedestrians are used for this calculation. On that way, the scenario size can be minimized and the flooding algorithm can be used in a fast manner

\begin{figure}
\centering
\includegraphics[width=0.6\linewidth]{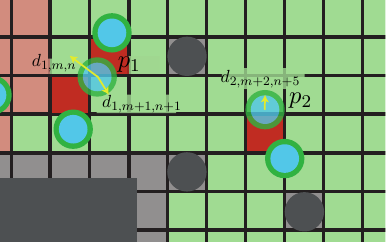}
\caption{Pedestrians and their corresponding transit-empty cells. The distance $d_{i,m,n}$ between a pedestrian $p_i$ and a cell $\bf{c_{m,n}}$ is shown by yellow arrows. Red marked are conflict cells between multiple pedestrians.}
\label{fig:FreeCells_o7}
\end{figure}

In the next step, we only consider spatial-continuous transformation candidates who intersect with at least one transit-empty cell $\bf{c_{m,n}}$ in their transit area. A cell $\bf{c_{m,n}}$ is transit-empty, if this cell neither contains an obstacle nor another pedestrian. If a spatial-continuous pedestrian has exactly one transit-empty cell, she will be transformed to this cell. If multiple transit-empty cells exist for a pedestrian, she will be transformed to the nearest transit-empty cell. The distance between a pedestrian $p_i$ and her corresponding cell ${c_{m,n}}$ can be calculated by the distance of their center-of-masses:
\begin{equation}
d_{i,m,n} =\left|\vec{p}_i -\vec{c}_{m,n}\right|
\end{equation}
A transformation means, that a pedestrian $p_i$ is removed from the SpaceCont model and transformed to the SpaceDisc model as a spatial-discrete pedestrian $p_j$. The new position is the center of the assigned cell $\bf{c_{m,n}}$:
\begin{equation}
\vec{p}_j = \vec{c}_{m,n} = \vec{p}_i + \vec{d}_{i,m,n}
\end{equation}
An example for this procedure is shown in Figure \ref{fig:FreeCells_o7}. Pedestrian $p_1$ has to decide between cell $\bf{c_{m,n}}$ and $\bf{c_{m+1,n+1}}$. The pedestrian will be transformed to cell  $\bf{c_{m+1,n+1}}$, due to the cell's smaller distance $d_{1,m+1,n+1}$. However,  $p_2$ is transformed to the cell $\bf{c_{m+2,n+5}}$, since this pedestrian intersects with only one transit-empty cell.

A local collision detection is necessary to check all intersecting cells $\bf{c_{m,n}}$ of a pedestrian $p_i$. In SpaceCont models, a pedestrian is typically represented by a circular or elliptic torso with radius $r_i$ and the geometry of cells are typically triangular, quadratic or hexagonal shaped (see Table \ref{tab2}). Since we follow a generic transition approach, our method has to be conservative. Otherwise, it would not be applicable for arbitrary pedestrian shapes and arbitrary cell geometries. Consequently, if we have elliptic shaped pedestrian torsos, we assume these as circular torsos with a radius $r_i$ equal to the size of the larger major axis. Different approaches exist to solve these 2D-collision detection problems \cite{lin1993cient, weller2013new}. 

After this procedure, all remaining transformation candidates are pedestrians, who intersect with non transit-empty cells. This means, these pedestrians intersect with cells, which are overlapped by more than one transformation candidate. It is important to note, that already occupied cells are inaccessible for a transformation. Occupied means, that these cells contain a spatial-discrete pedestrian, intersect with an obstacle or intersect with spatial-continuous pedestrians who are no transformation candidates. 

\begin{figure}
	\centering
	\includegraphics[width=0.6\linewidth]{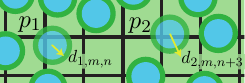}
	\caption{Competing pedestrians and the corresponding available grid cells. The shortest distance $d_{i,m,n}$ of a pedestrian $p_i$ for a cell $\bf{c_{m,n}}$ is shown by yellow arrows.}
	\label{fig:Concurent_Peds04}
\end{figure}

Since the remaining transformation candidates do not have any transit-empty cells, they have to compete with other transformation candidates. The problem can be resolved by assigning the pedestrian with the lowest center-of-mass distance $d_{i,m.n}$ to a cell $\bf{c_{m,n}}$. If multiple pedestrians intersect with one cell, the nearest pedestrian is assigned to this cell. If a pedestrian has the lowest center-of-mass distance to multiple cells, the pedestrian is assigned to the nearest of these cells. This procedure can be seen in Figure \ref{fig:Concurent_Peds04}. Pedestrian $p_1$ and $p_2$ are the agents with the lowest distance $d_{1,m,n}$ respectively $d_{2,m,n+3}$ for the cells $\bf{c_{m,n}}$ respectively $\bf{c_{m,n+3}}$.

\begin{figure}
	\centering
	\includegraphics[width=0.6\linewidth]{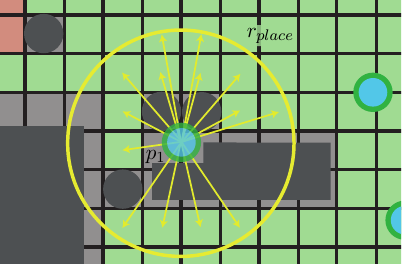}
	\caption{Available grid cells for a pedestrian $p_i$ in the maximal distance $r_{place}$ (yellow circle)}
	\label{fig:r_place03}
\end{figure}

After performing these steps, there may remain pedestrians $p_i$ that could not be assigned to one of the cells $\bf{c_{m,n}}$ they are intersecting with. Consequently, these agents have to be transformed to neighbouring or even more distant cells. However, to transform such a pedestrian, we need to find all surrounding free and therefore available cells in a distance $r_{place}$. The radius $r_{place}$ is chosen in such a way, that the errors caused by the spatial transformation into a more distant cell is not larger than the spatial error caused by not transforming the pedestrian at all during this time step. 
Since the pedestrian $p_i$ has a maximal velocity about $v_{max}$ and the next transformation process takes place after a simulation time step $\Delta t_{disc}$, the maximal spatial error caused by the missing transformation is:
\begin{equation}
r_{place} = v_{max} \cdot \Delta t_{disc}
\label{EQ:rplace}
\end{equation}
All free cells in a distance radius $r_{place}$ from the pedestrian position $\vec{p}_i$ are considered as possible transformation cells. In each transit area, the possible transformation cells are determined for all remaining spatial-continuous pedestrians $p_i$. The pedestrian in a transit area with the lowest number of free available cells is assigned to her nearest free cell. Afterwards, this cell is blocked for all other remaining pedestrians of this transit area. This procedure is repeated until all pedestrians are assigned to a free cell. An example is shown in Figure \ref{fig:r_place03}. Different cells are available for pedestrian $p_i$ (yellow arrows). A cell can be closed for the transformation due to two different reasons: it can be blocked either because of an obstacle or a spatial-discrete pedestrian. It is important to remember, that all spatial-continuous pedestrians, who were transformed in the previous steps of the whole transformation process have to be considered as spatial-discrete pedestrians located at their assigned cell positions. 

In Table \ref{tab2} we calculated the largest possible density for different simulation model types. However, even at lower local densities difficulties may occur for the generic transformation procedure. This happens if the transit area gets too crowded. In this case, no free spaces are available for the transformation of pedestrians. Consequently, the transformation process can not be executed. A solution for this issue is the use of a dynamic zoom approach, which we describe in Section \ref{sec:DynamicZoom}.   

 \subsection{Transformation from the SpaceDisc model to the SpaceCont model}
 \label{sec:TraMe2Mi}

In the following, we describe the transformation process from spatial-discrete to spatial-continuous. This transformation direction can be executed quite easily, since each spatial-discrete pedestrian has sufficiently enough space for the transformation to the spatial-continuous scale. Again, only pedestrian located in a transit area can be used for this process. At first, a propagation vector has to be calculated:
\begin{equation}
\vec{\sigma}_j = \vec{e}_{v_j}\cdot v_{max}\cdot \Delta t_{disc}
\end{equation}
Furthermore, the same definition of the surrounding angle $\pm\Omega_j$ can be used for the specification of the corresponding propagation segment of a spatial-discrete pedestrian $p_j$:
\begin{equation}
\pm\Omega_j = \pm\frac{\Delta t_{disc}\cdot v_j}{d_s(v_j)}\cdot |\omega_s|
\end{equation}
\begin{figure}
\centering
\includegraphics[width=0.6\linewidth]{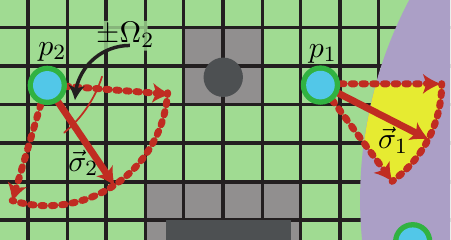}
\caption{Propagation vector $\vec{\sigma}_j$ and the surrounding angle $\pm\Omega_j$ define shape and size of the propagation segment (yellow marked area)}
\label{fig:MesoMicro_Propagation__03}
\end{figure}
Figure \ref{fig:MesoMicro_Propagation__03} shows an exemplary use of the propagation segment. Pedestrian's $p_1$ segment intersects with a spatial-continuous area and therefore this pedestrian has to be transformed to the SpaceCont model. The propagation segment of pedestrian $p_2$ contains the transit area only. Consequently, this pedestrian will not be transformed. 

This process describes the transformation of a low-resolution pedestrian $p_j$ to a high resolution pedestrian $p_i$. A spatial-discrete pedestrian $p_j$ is always located on a cell $\bf{c_{m,n}}$. If no spatial-continuous pedestrian $p_i$ is placed inside this cell $\bf{c_{m,n}}$, $p_j$ can be transformed to on the center position of her current cell:
\begin{equation}
\vec{p}_j = \vec{c}_{m,n} = \vec{p}_i
\end{equation} 

The velocity is $v_j$ of the transformed pedestrian equals the former velocity $v_i$ (see Section \ref{sec:TransVelocity}). If virtual pedestrians are used for the transit areas (see Section \ref{sec:virtualPeds}), the cell of a spatial-discrete pedestrian can not intersect with any other obstacle or pedestrian. Consequently, the pedestrian $p_j$ can be transformed to the SpaceCont model on the center position of this cell. 

\subsection{Velocity Transformation}
\label{sec:TransVelocity}
The transformation of velocities from transformed pedestrians is executed via the parameter interface (see Table \ref{rat}). Every time step the \TransiTUM{} framework receives the current velocities of all pedestrians from its SpaceDisc and SpaceCont models. If a pedestrian is transformed to a new model, the former velocity is given via this interface to the new model. Afterwards, it is up the model itself how it uses the velocity information.

In the interface itself the velocity is saved as a float number. In SpaceCont models the velocity in the interface can be used directly most of the time, since these models normally use float numbers. However, SpaceCont models have a limited spatial resolution. Therefore, the velocity has to be scaled down. Since the pedestrians of SpaceDisc models move cell wise, the numerical velocity has to be translated to cells per time step:

\begin{equation}
\frac{\mbox{number of cells}\cdot \mbox{cell size}}{\Delta t_{disc}} =\vec{v}_i
\end{equation}

Since in SpaceDisc models the pedestrian can be only moved by a discrete number of cells, the velocity can not be exactly transformed from SpaceCont to SpaceDisc. The way this is dealt with depends on the underlying model and has to be implemented in the interface between SpaceDisc model and the \TransiTUM{} framework.

\subsection{Limitations of the generic transition}
\label{sec:Limit}
The \TransiTUM{} framework is a comprehensive and capable method to couple arbitrary pedestrian dynamics models. However, due to its generic approach some limitations exist. Since both pedestrian dynamics models run independently from each other, one major concern is the overlapping of spatial-continuous and spatial-discrete pedestrians in the transit area. However, the use of virtual pedestrians (see Section \ref{sec:virtualPeds}) in the transit area solves this issue. Another issue is the transformation of model individual attributes, e.g. group behavior \cite{moussaid2010walking}. Since such attributes are quite model specific, they can not be transferred in a generic approach. Consequently, such information gets lost during the transformation process. Furthermore, inaccuracies can occur in the transit area. If pedestrians are placed on new positions during the transformation process, the intrinsic simulation model is disturbed by these transformation artifacts. Therefore, results from the transit areas should be considered with particular caution. Nevertheless, the largest concern of a generic transition approach are very high pedestrian densities in a transit area. If the pedestrian density gets too high, it is possible that a pedestrian cannot be transformed although she could leave the transit area during the next spatial-discrete time step. Consequently, it is not possible to execute the corresponding transformation.

However, a dynamic Zoom-In approach can be used to solve this problem, since this method prevents the transit area from getting too overcrowded.

\section{Dynamic Zoom-In approach}
\label{sec:DynamicZoom}
\subsection{Dynamic transformation scenarios}
\label{deTrSec:31}
The dynamic Zoom-In approach extends our generic transition by a dynamic component to consider time-dependent changes of pedestrian densities during a simulation  (see Figure \ref{fig:overview03}).

\begin{figure}
	\centering
	\includegraphics[width=1\linewidth]{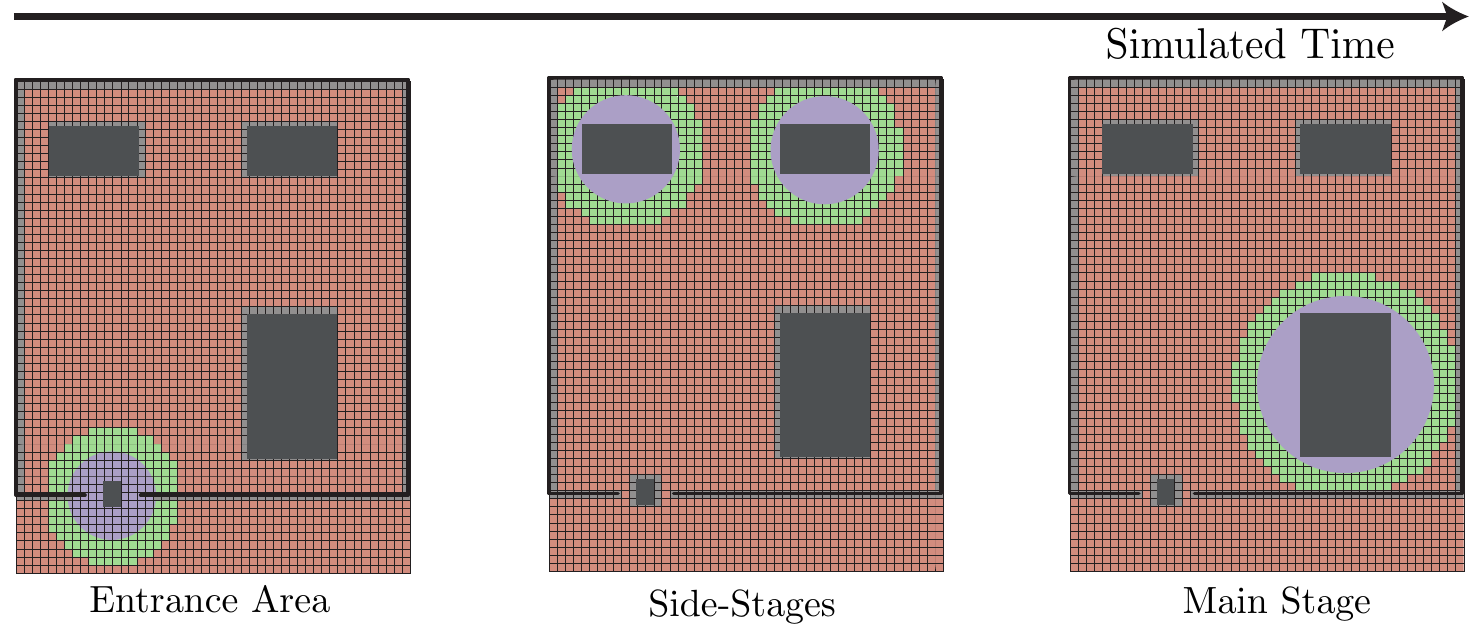}
	\caption{Timely procedure of the Zoom-In approach for a festival scenario}
	\label{fig:FestivalExample_15}
\end{figure}

Regions with critical situations can change during the timely progress of a simulation. Thus, an automatic detection of regions with potentially dangerous densities is necessary. A good example is the simulation of a music festival with an entrance area, two side-stages and one main stage (see Figure \ref{fig:FestivalExample_15}). At the beginning of the event, all visitors are flooding in through the entrance area. Consequently, this area is the most sensitive region and has to be simulated in detail. During the pre-program, the opening acts play on the two side-stages. Most of the visitors have entered the event site and will stay next to one of these stages to enjoy the music. Therefore, the entrance area has lost relevance and can now be simulated by the less computational costly SpaceDisc model. In exchange, the side-stages have to be calculated by the SpaceCont model due to the increasing local densities in these areas. After some time, the main act starts and attracts the visitors' attention. So the crowd flow moves from the side-stages to the main stage. From this moment on, the main stage has to be simulated in full detail, while the side-stages can be calculated by a SpaceDisc model. Such dynamic switches enable the optimal use of available computational resources. 

\subsection{Density determination}
\label{deIFF:32}

The determination of densities in the field of pedestrian dynamics is  a complex process and was well-studied in different research projects (e.g. \cite{steffen2010methods, Zhang2014, duives2015quantification}). The main issue to calculate density distributions of pedestrians is their size. The size of a pedestrian ($\approx\SI{1}{\metre}$) is typically the same magnitude as the spatial resolution we would like to observe ($\approx\SI{1}{\metre}$). Therefore, a standard static method like counting pedestrians on an observed area will suffer from ''large data scatter'' \cite{steffen2010methods} due to the low number of pedestrians on such an area. Thus, it is difficult to obtain locally a smooth density distribution. Different methods exist to solve this issue \cite{duives2015quantification}. The most common one is the Voronoi-method \cite{Zhang2014}. The Voronoi-method assigns to each pedestrian in the observed area a Voronoi-cell. This Voronoi-cell contains all points  which are closer to the position of a pedestrian $p_i$ than to any other pedestrian. 
This gives a much smoother density distribution than with the standard static calculation method. 

Another well-known approach is the so called XT-method \cite{duives2015quantification,ni2004direct,edie1963discussion}. This method counts the number of pedestrians per grid cells over a given time interval. The density per cell is calculated by the averaged number of pedestrians, who visited this cell during the considered period of time $\Delta T$. For each cell, the pedestrians are weighted by their length of stay. Consequently, the time dependent density $\rho_{m,n}(t)$ for a cell $\bf{c_{m,n}}$ can be determined by the following equation (based on \cite{duives2015quantification} and applied to discrete simulation time steps):
\begin{equation}
\rho_{m,n}(t) = \frac{\sum_i k_{i,c_{m,n}}(t)\cdot\Delta t_{cont} + \sum_j k_{j,c_{m,n}}(t)\cdot\Delta t_{disc}}{A_{m,n}\cdot\Delta T}
\end{equation}

A pedestrian is only counted if her position $\vec{p}_i$ respectively $\vec{p}_j$ is part of the cell's area $A_{m,n}$. Parameter  $k_{j,c_{m,n}}$ represents the number of spatial-discrete time steps that the pedestrian was located inside of cell $\bf{c_{m,n}}$ during the measured time frame $\Delta T$: 
\begin{equation}
k_{j,c_{m,n}}(t) = \sum_{n=M}^{N} \delta_{j,c_{m,n}}(t_n)
\end{equation}
This sum iterates over all spatial-discrete time steps $t_n$, which are part of the measured time interval $[t-\Delta T, t]$. Consequently, the boundary indexes are $M = \left\lceil\frac{t-\Delta T}{\Delta t_{disc}}\right\rceil$ and $N=\left\lfloor\frac{t}{\Delta t_{disc}}\right\rfloor$. The cell-based Dirac measure $\delta_{j,c_{m,n}}(t_n)$ equals one if the pedestrian is part of the cell $\bf{c_{m,n}}$ at the time step $t_n$ and zero in any other case:
\begin{equation}
\delta_{j,c_{m,n}}(t_n)= \begin{cases} 0, & \vec{p}_j(t_n)\notin\bf{c_{m,n}} \\ 1, & \vec{p}_j(t_n)\in\bf{c_{m,n}} \end{cases}
\end{equation}
The parameter $k_{i,c_{m,n}}(t)$ for the number of spatial-continuous time steps can be calculated in a similar way:
\begin{equation}
k_{i,c_{m,n}}(t) = \sum_{n=M}^{N+1}\sum_{l=L_\downarrow}^{L_\uparrow} \delta_{i,c_{m,n}}(\tau_{n,l})
\end{equation}
The indexes for the first sum iterate from the spatial-discrete time step before the measured interval $[t-\Delta T, t]$ starts to the spatial-discrete time step after this interval ends. Consequently, the boundary indexes are defined as $\left\lfloor\frac{t-\Delta T}{\Delta t_{disc}}\right\rfloor = M-1$ and $\left\lceil\frac{t}{\Delta t_{disc}}\right\rceil = N+1$. Our indexes $L_\downarrow$ and $L_\uparrow$ for the second sum ensure that only spatial-continuous time steps are considered, which are part of the measured time interval. Therefore, the starting index $L_\downarrow$ is given by the following equation:
\begin{equation}
L_\downarrow= \begin{cases} \left\lceil \frac{(t-\Delta T)-\tau_{M-1,d_{M-1}}}{\Delta t_{cont}}\right\rceil, & t - \Delta T \geq t_n \\ \qquad\qquad\qquad\quad\quad\,\,\,\, 1, & t - \Delta T < t_n \end{cases}
\end{equation}
The ending index $L_\uparrow$ for the second sum is defined as:
\begin{equation}
L_\uparrow= \begin{cases} d_n - \left\lfloor \frac{\tau_{N+2,1}- t}{\Delta t_{cont}}\right\rfloor, & t  \leq t_n \\ \qquad\qquad\qquad d_n, & t  > t_n \end{cases}
\end{equation}
Since we have a dynamic problem, the XT-method is more sufficient than the classic Voronoi approach. Consequently, we use the XT-method to calculate the local densities for our Zoom-In approach.  

\subsection{Dynamic Zoom-In: Switching from SpaceDisc to the SpaceCont model}
\label{sec:ZoomIn}
Based on the measured densities, we determine the areas, which have to be switched from the SpaceDisc model to the spatial-continuous one. To this end, we have to introduce a parameter $\rho_{thr}$, which describes the maximal tolerable density value before a SpaceCont simulation is necessary. SpaceDisc models have intrinsically restricted maximal densities according to their cell geometries. An overview of the limited maximal density for different cell geometries was described in Table \ref{tab2}. However, the walking behavior of pedestrians is affected already by significantly lower densities. According to Weidmann \cite{weidmann1992transporttechnik}, the velocity of a pedestrian is halved at a local density of $\SI{1.5}{ped\per\meter\squared}$. Therefore, we recommend $\rho_{thr}=\SI{1.5}{ped\per\meter\squared}$ as a limit for the maximal tolerable density for the use of SpaceDisc models. Furthermore, we need to define an influence radius $R$. If a cell is chosen as the center for a spatial-continuous circular area, all cells with a distance smaller than $R$ are part of its influence sphere. The size of $R$ can be described by the following equation:
\begin{equation}
R \leq r_{place} =\Delta t_{disc} \cdot v_{max}
\end{equation} 
This definition is reasonable for SpaceDisc models, since this value describes the maximal distance a spatial-discrete pedestrian can move during her next time step.

After the definition of these parameters, we are able to define the Zoom-In checking procedure, which determines areas with critical high densities:
\begin{enumerate}
	\item All cells $\bf{c_{m,n}}$ are ordered according to their XT-density during the last measuring interval $[t-\Delta t, t]$.
	\item The cell $\bf{c^*}$ with the largest XT-density $\rho^*$  is identified.
	\item A possible transformation area is found, if the cells density $\rho^*_{m,n}$ extends the density threshold
	\begin{equation}
	\rho^* \geq \rho_{thr}
	\end{equation}
	\item All cells $\bf{c_{m,n}}$ in a distance of $\left| \vec{c}^*-\vec{c}_{m,n}\right|\leq R$  are in the sphere of influence from $\bf{c^*}$. These cells are removed from the order in Step 1. 
	\item Beginning at Step 2, the procedure is repeated until the new maximal density falls below the threshold:
	\begin{equation}
\rho^* < \rho_{thr}
	\end{equation}
\end{enumerate} 
After the Zoom-In checking procedure, we have identified a set of cells, which are possible Zoom-In candidates $Z_\phi=\left\{ {c}^*|\rho^*\geq\rho_{thr}\right\}$. In the next step, we calculate the surrounding local density to determine the necessary size of the spatial-continuous area. For each candidate $\bf{c^*}$ only cells within a given maximal distance of $k\cdot R$ are considered for the local density $\rho_{loc}$. This can be described by the Heaviside step function $\mathcal{H}(x)$:
\begin{equation}
\mathcal{H}(x) = \frac{d}{dx}\max\left\{x,0\right\}=\\\left\{\begin{array}{ccl}0, && x< 0 \\ 1, && x\geq 0\end{array}
\right.
\end{equation}
Furthermore, all cells which contain obstacles are excluded from our calculations, since these cells can not contain any pedestrians:
\begin{equation}
\delta^*_{c_{m,n}}= \begin{cases} 0, & \mbox{obstacle}\in\bf{c_{m,n}} \\ 1, & \mbox{obstacle}\notin\bf{c_{m,n}} \end{cases}
\end{equation} 
Consequently, a selection function can be defined to determine if a cell $\bf{c_{m,n}}$ is obstacle free and within the maximal distance $k\cdot R$:
\begin{equation}
\alpha(c_{m,n},\vec{c}^*,k) = \mathcal{H}(k\cdot R-|\vec{c}^*-\vec{c}_{m,n}|)\cdot\delta^*_{c_{m,n}}
\end{equation}
Based on the selection function, the average time depending local density $\rho_{loc}(\vec{c}^*,t,k)$ of the cell  $\bf{c^*}$ is calculated by:
\begin{equation}
\rho_{loc}(\vec{c}^*,t,k) = \frac{\sum_{m,n} \rho_{m,n}(t)\cdot\alpha(c_{m,n},\vec{c}^*,k)}{\sum_{m,n}\alpha(c_{m,n},\vec{c}^*,k)}
\label{eq:SizeOfMicriDings}
\end{equation}
We use the following procedure to determine the necessary size of the spatial-continuous area:
\begin{enumerate}
	\item The candidate $\bf{c^*}\in Z_\phi$ with the largest XT-density $\rho^*$  is identified with $k=1$.
	\item All cells $\bf{c_{\#}^*}\in Z_\phi$ in the spatial-continuous zone of $\bf{c^*}$ ($|\vec{c}^*-\vec{c}^*_\#|\leq k\cdot R$) are removed as possible candidates for further spatial-continuous zones.
	\item The local density $\rho_{loc}(\vec{c}^*,t,k)$ is calculated for $\bf{c^*}$.
	\item If the local density extends the density threshold ($\rho_{loc}(\vec{c}^*,t,k)\geq\rho_{thr}$), $k$ is increased by one ($k\rightarrow k+1$) and the procedure repeated by Step 2.
	\item The algorithm stops by $k=k_{max}$ and starts with Step 1 for the next $\bf{c^*}$.
\end{enumerate}
After the algorithm terminated, we need to define the center $\vec{s}_{CoM}$ of the spatial-continuous zone. The most dense region has to be as close as possible to the spatial-continuous center. Therefore, the position of the densities' center-of-mass should be used as the center of this spatial-continuous area:
\begin{equation}
\vec{s}_{CoM}(\vec{c}^*,t,k) = \frac{\sum_{m,n}\rho_{m,n}(t)\cdot\alpha(c_{m,n},\vec{c}^*,k)\cdot\vec{c}_{m,n}}{\sum_{m,n}\rho_{m,n}(t)\cdot\alpha(c_{m,n},\vec{c}^*,k)}
\end{equation}
\begin{figure}
	\centering
	\includegraphics[width=0.8\linewidth]{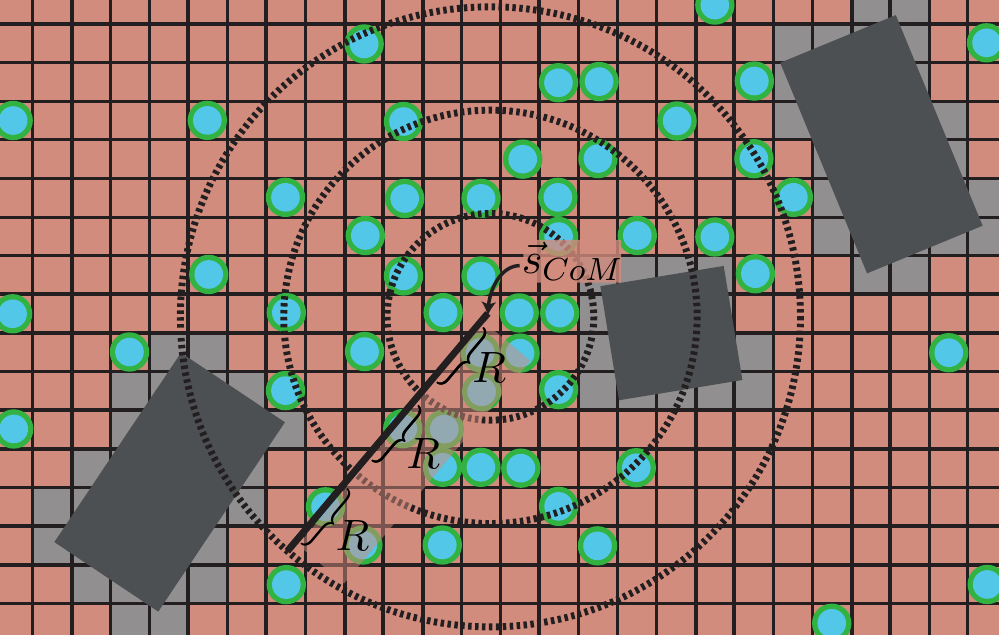}
	\caption{Circular shaped spatial-continuous area determined by the Zoom-In approach}
	\label{fig:MicroZones_06}
\end{figure}
The spatial-continuous area for a candidate $\bf{c^*}$ has a circular shape with radius $k\cdot R$ and a center at $\vec{s}_{CoM}$. Figure \ref{fig:MicroZones_06} shows an exemplary spatial-continuous zoom area with radius $k\cdot R=3R$. 

In the next step, all spatial-discrete pedestrians within the determined spatial-continuous area have to be transformed to the SpaceCont model. This procedure follows the process described at Section \ref{sec:TraMe2Mi} with the difference that all pedestrians $p_j$ in this area are transformed according to their current position $\vec{p}_j$. This means that no propagated position calculations are necessary. Furthermore, the spatial-continuous area has to be surrounded by a transit area with width $w_{Tr} > v_{max}\cdot\Delta t_{disc}$ (see Section \ref{subsec:TheTransiZOnes}).  

\subsection{Dynamic Zoom-Out: Switching from the SpaceCont model to the SpaceDisc model}
\label{sec:ZoomOut}
The Zoom-Out procedure is necessary to save computational time. If an area gets less crowded during the simulation run, it is possible to change the underlying model from SpaceCont to SpaceDisc. This process works similar to the Zoom-In approach described in the previous Section \ref{sec:ZoomIn}. We determine the current local density for a spatial-continuous area starting at the outer ring. The local density $\rho_{\circ}(\vec{s}_{CoM},t,k)$ for the $k^{\text{\tiny th}}$ ring of the spatial-continuous zone can be calculated as:
\begin{equation}
\rho_{\circ}(\vec{s}_{CoM},t,k) = \frac{\rho_{\circ}(\vec{s}_{CoM},t,k)\cdot k^2 - \rho_{\circ}(\vec{s}_{CoM},t,k-1)\cdot (k-1)^2}{k^2 - (k-1)^2}
\end{equation}
Each spatial-continuous area with its center at $\vec{s}_{CoM}$ and its radius $k\cdot R$ has to undergo the following procedure: 
\begin{enumerate}
	\item The average density $\rho_{\circ}(\vec{s}_{CoM},t,k)$ is calculated for all cells $\bf{c_{m,n}}$ in the outer ring of the spatial-continuous area: 
	\begin{equation}
	(k-1)\cdot R\leq |\vec{s}_{CoM}-\vec{c}_{m,n}|\leq k\cdot R
	\end{equation}
	\item The algorithm stops, if the average density $\rho_{\circ}(\vec{s}_{CoM},t,k)$ extends the threshold density, since the density is too high to switch back to the SpaceDisc model:	
		\begin{equation}
	\rho_{\circ}(\vec{s}_{CoM},t,k) \geq \rho_{thr}
		\end{equation}
	\item If parameter $k$ is not smaller than two ($k\geq2$), this parameter gets lowered by one ($k\rightarrow k-1$), which means that the next smaller ring is investigated. The procedure is repeated at Step 1. 
\end{enumerate}
The remaining size of the spatial-continuous area is given by the remaining size of $k$. If parameter $k$ was decreased to zero ($k=0$) during the described procedure, the whole spatial-continuous area is transformed to the spatial-discrete scale. If the parameter is larger ($k>0$), only the outer rings are transformed to the SpaceDisc model, while the inner rings stay spatial-continuous. The transformation process follows the description in Section \ref{sec:TranMi2Me}. However, no propagation calculations are necessary since all pedestrians $p_i$ in the spatial-continuous area are transformed according to their current position $\vec{p}_i$. A new transformation zone with $\omega_{Tr}$ is defined around the reduced spatial-continuous area.

\section{Proof of concept}
\label{cdtfasd4}
\subsection{The simulation scenario}
\label{cdtfasd4o1}
\begin{figure}
	\centering
	\includegraphics[width=\linewidth]{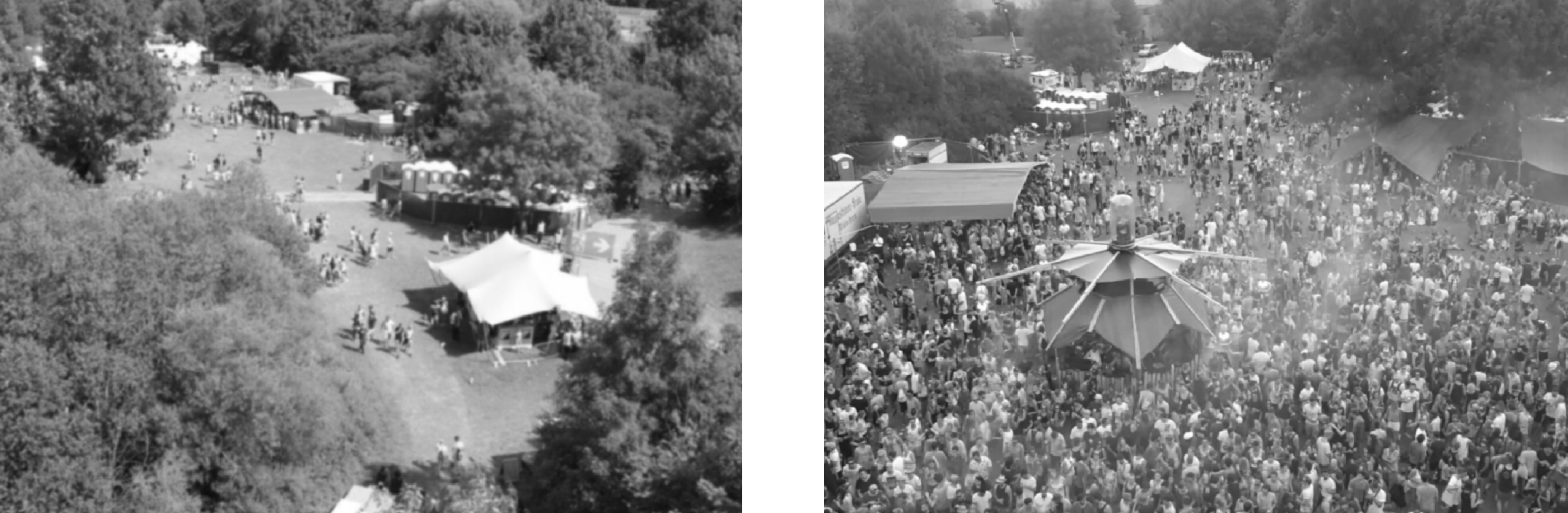}
	\caption{The event site at the beginning of the festival and after most visitors entered the location (taken by Michael \"{O}hlhorn, Vabeg Eventsafety)}
	\label{fig:BildBTTW_03}
\end{figure}
Our simulation scenario is an annual local music festival with approximately 5000 visitors (see Figure \ref{fig:BildBTTW_03}). Our research team observed this event over three consecutive years from 2014 to 2016. The age structure of the audience is quite homogeneous, since most of the visitors are young adults, mostly students, under the age of 30 years. The event site is located in the surrounding area of Munich (Germany) at the campus of the Technical University of Munich. The event site itself is located at a remote area of the campus and surrounded by woodland. This open-air festival is a one day event and takes place at the end of July. The largest share of incoming visitors arrives with the local public transport services at a near subway station. From this place, the visitors walk to the actual event site.

\begin{figure}
\centering
\includegraphics[width=1\linewidth]{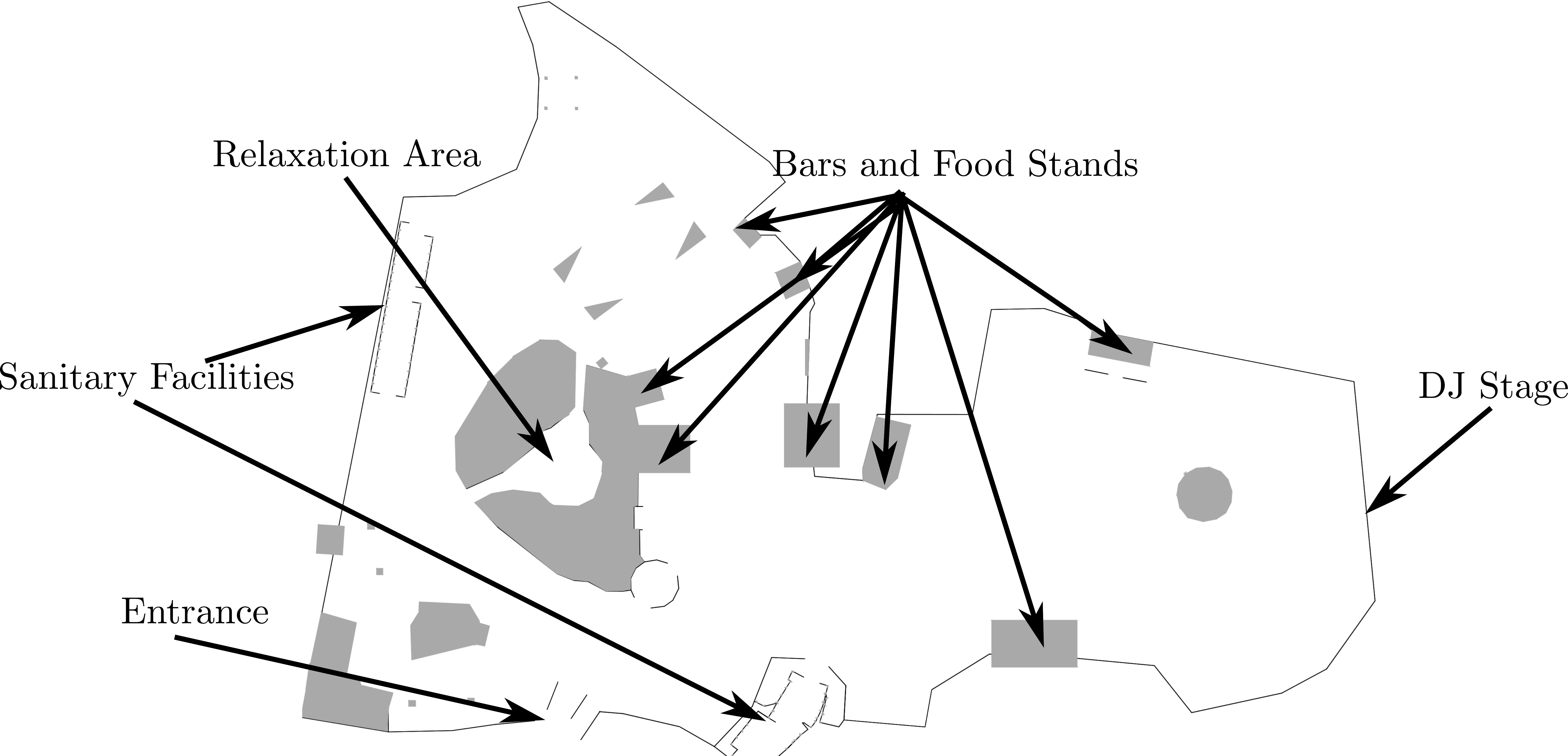}
\caption{Simulation scenario from the annual music festival in the summer of 2016}
\label{fig:scenario_04}
\end{figure}

The layout of the event site from 2016 can be seen in Figure \ref{fig:scenario_04}. Its total diameter was about $\SI{200}{\meter}$ in the length and $\SI{100}{\meter}$ in the width. Its entrance was located at the south and the mobile sanitary facilities were spread at the outside edges of the festival. 
The main attraction was the DJ stage at the east as well as the dancing area in front of it. The right picture of Figure \ref{fig:BildBTTW_03} shows this dancing area at a peak hour. This layout was created via common computer aided design software and used for the simulation. A desired velocity of $\SI{1.34}{\meter\per\second}$ was used for the visitors on the event site \cite{weidmann1992transporttechnik}. More detailed information about the overall boundary conditions of this annual event can be found by Ehrecke \cite{Ehrecke:2016:Bachelorthesis}.

\subsection{Used simulation layers}
\label{cdtfasd4o2}
\begin{figure}
	\centering
	\includegraphics[width=1.0\linewidth]{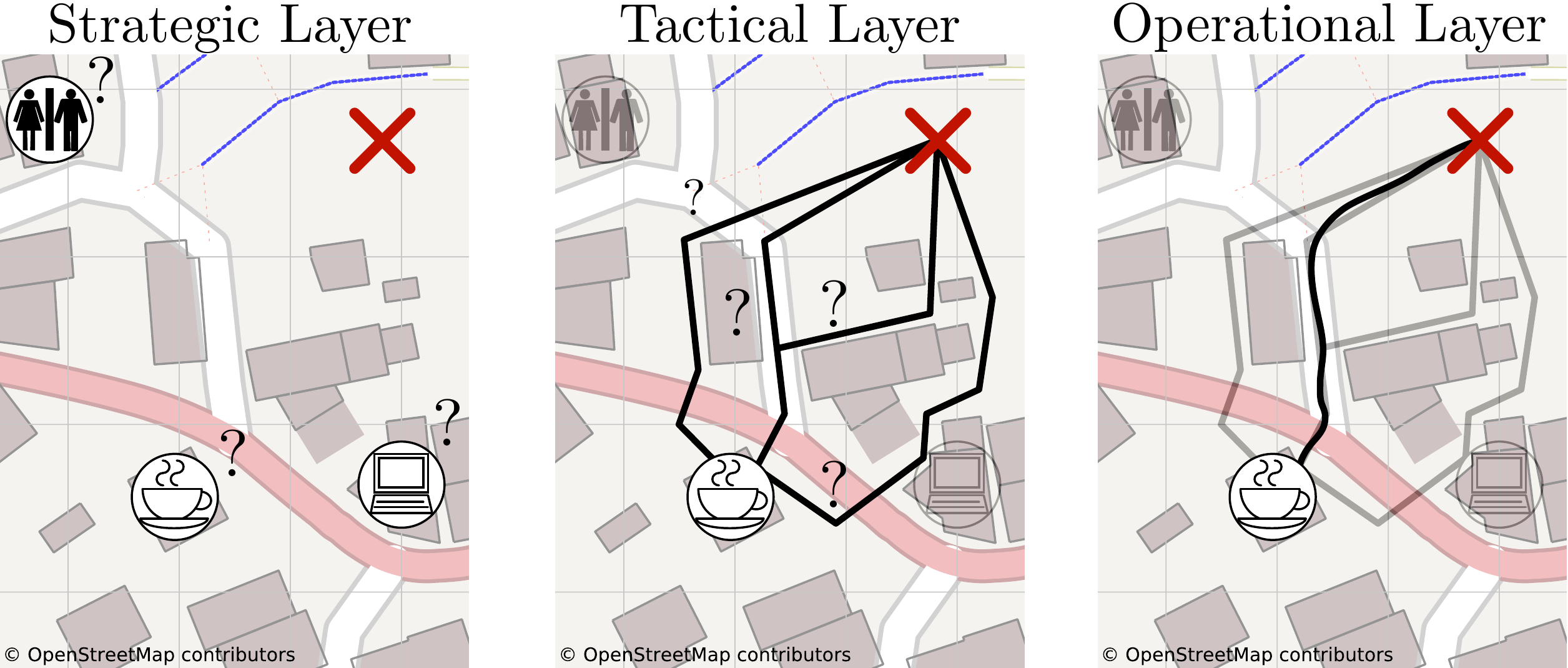}
	\caption{The three layer approach from Hogendoorn \cite{hoogendoorn2002microscopic}: the strategic target selection, the tactical routing selection and the operational locomotion of pedestrians.}
	\label{fig:3LayerHoogendorn}
\end{figure}

Our concepts were implemented in a software prototype to validate their applicability as a proof of concept (see Figure \ref{fig:overview03}). The implementation is based on the three layer approach (strategic, tactical, operational) from Hogendoorn \cite{hoogendoorn2002microscopic} and was integrated into the \MomenTUM{} simulation framework \cite{MOMENTUM}. The strategic layer determines the spatial target of a pedestrian, the tactical layer calculates the route to this target and the operational layer executes the movement of pedestrian (See Figure \ref{fig:3LayerHoogendorn}).

For the strategic layer, a well-established method is the classical origin-destination (OD) matrix approach. The OD-matrix assigns to each origin-destination pair a probability that a pedestrian, generated in this origin, will choose the associated destination as their strategic target. On that way, a randomly assigned sequence of visiting points on the event site can be applied to each pedestrian. However, for such a complex scenario with many different targets, a more realistic method like the cognitive science based approach from Kielar and Borrmann \cite{Kielar201647} is useful. Their model is psychological inspired and uses an interest function to describe the destination choice of pedestrians.

Different models exist for the tactical layer, e.g. cognitive maps, which restrict the spatial knowledge of pedestrians in a realistic way \cite{10.1007/978-3-319-33482-0_32}. Other researchers gather empirical information to describe the decision process of a pedestrian on the tactical layer \cite{articletest}. However, the event site of the simulated festival is quite open and all possible targets are easy to reach. All possible routes between the different targets were calculated by a static visibility graph over the whole scenario layout \cite{MOMENTUM}. Since we simulate the course of an event and not an evacuation situation, a shortest path approach \cite{dijkstra1959note} is not sufficient to calculate the route to the next target. Consequently, we used the unified routing model to apply a more realistic rout choice behavior \cite{kielar2015unifiedRouting} on this visibility graph.

The operational layer has to be simulated by two different models, a spatial-continuous and a spatial-discrete one. The operational layer moves the pedestrian along the visibility graph from routing point to routing point to the desired target. For the spatial-continuous model, the Moussa{\"\i}d and Nelson heuristic model, a Social-Force variant, was used \cite{moussaid2014simple, PhysRevE.87.063305}. 
For this model, a relaxation time of $\SI{0.5}{\second}$ was used for the calculation of the acceleration. The mass behavior factors for the repulsive interaction force were $\SI{26.67}{\kilogram\meter\per\square\second}$ and $\SI{0.06}{\meter}$. Furthermore, the sliding friction force factors were $\kappa=\SI{2.4e5}{\kilogram\per\square\second}$ and $k=\SI{1.2e5}{\kilogram\per\meter\per\second}$.

The spatial-discrete part of the operational layer was simulated by the ''Cellular Stock'' model developed by our group \cite{Biedermann:2016:MobilTUM, ThesisDHBiedermann}. It is partially based on the operational approaches used by Bandini et al. \cite{Bandini2011} as well as Blue and Adler \cite{Blue2001}. A grid of quadratic cells with an edge size of $\SI{0.46}{\meter}$ was used as the underlying lattice for this cellular automaton.

This cellular automaton model can be used for quadratic and hexagonal grid cells. The distance a pedestrian $p_j$ has to walk between two cells equals the walking distance $\omega$ between the center of these two cells. Based on this distance, the model calculates a walking stock to determine if and in which order a pedestrian is moved during the spatial-discrete simulation time steps. The algorithm works as follows for each pedestrian $p_j$ for each simulated time step:

\begin{itemize}
	\item Increase the walking stock $S_j$ by $v_{des}\cdot\Delta t$ with $v_{des}$ as the desired velocity and $\Delta t$ as the size of a time step.
	\item Determine all free neighboring cells, which are closer to the next routing point than the current cell center $\vec{c}_j$
	\item From these cells, chose the cell $c_b$ closest to the beeline between the previous and the next routing point 
	\item If $S_j\geq \left| \vec{c}_j - \vec{c}_b\right|$ move the pedestrian to $c_b$ and lower the stock by $\left| \vec{c}_j - \vec{c}_b\right|$
	\item If $p_j$ was not moved and $S_j > k\cdot v_des\cdot\Delta t$ with $k>1$, choose a random free neighboring cell $c_r$
	\item Move the pedestrian to cell $c_r$ and lower the stock by $\left| \vec{c}_j - \vec{c}_r\right|$
\end{itemize}

However, the three standard layers from Hogendoorn \cite{hoogendoorn2002microscopic} are not sufficient for a holistic and multi-scale simulation of an event. Kielar and Borrmann \cite{Kielar:2016:PED} proposed a set of spatial task solving models, which are necessary to enable the simulation of complex pedestrian behavior. In the case of a public event simulation, one important additional spatial task solving model is queuing in front of the entrance or vendor stands. The used simulator framework \MomenTUM{} supports such behavior models \cite{MOMENTUM}.

For the \TransiTUM{} approach we used a minimal radius of $\SI{2.0}{\meter}$ with a tolerable maximal density of $\SI{0.8}{ped\per\meter\tothe{2}}$. This means, that a SpaceCont model is applied as soon as the density in a circular area of $\SI{2.0}{\meter}$ extends $\SI{0.8}{ped\per\meter\tothe{2}}$. The density for the scenario and therefore the current spatial-continuous areas were calculated every two simulated seconds.

\subsection{Simulation procedure}

\begin{figure}
\centering
\includegraphics[width=1\linewidth]{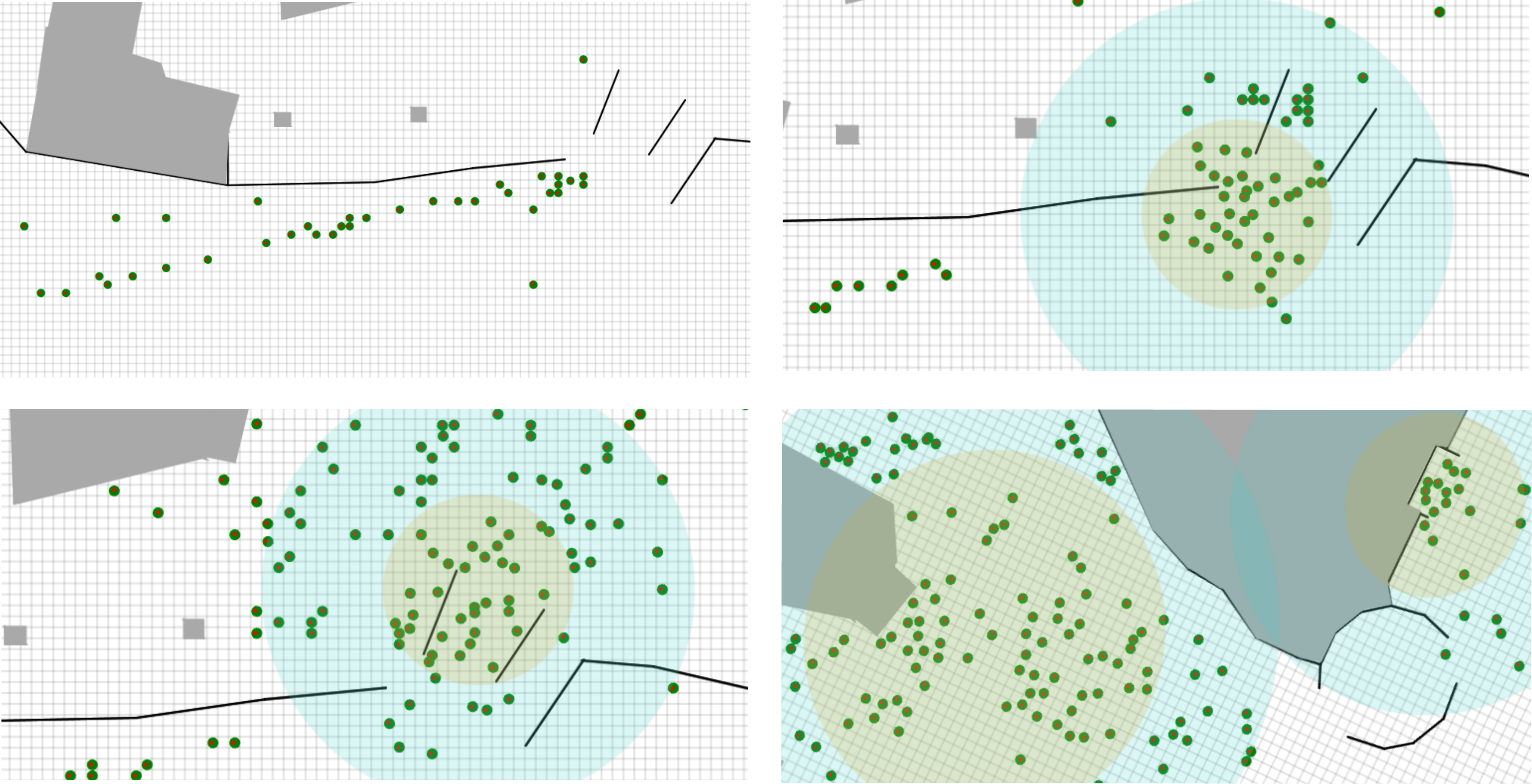}
\caption{Screenshots from the dynamic bidirectional coupling procedure during the simulation run.}
\label{fig:4FiduresSimulation_04}
\end{figure}

 During the simulation run the spatial-continuous areas move their location and adjust according to the current local pedestrian density. Figure \ref{fig:4FiduresSimulation_04} shows situations from the simulated music festival at different time steps. Areas of the SpaceCont model are colored in a slight ochre, the transit area in a slight mint and the sphere of the SpaceDisc model is uncolored. Spatial-discrete pedestrians move cell-by-cell throughout the scenario. Consequently, these pedestrians are placed perfectly in the center of their current grid cell. However, spatial-continuous pedestrians move continuously and thus they would mainly intersect with the cell edges of the grid. These screenshots show the bidirectional transformation procedure for pedestrians heading towards the same target: from the SpaceDisc to the SpaceCont model and from the spatial-continuous to the spatial-discrete one. Pedestrians in the spatial-discrete area are placed in the center of the grid cells, while the spatial-continuous pedestrians show a continuous movement. However, both cases occur in the transit area. At first, the pedestrians are entering the festival event area. Since only a small number of pedestrians has entered the scenario, the SpaceDisc approach is sufficient and no Zoom-In is necessary. This is shown in the first screenshot of Figure \ref{fig:4FiduresSimulation_04}. As more and more pedestrians are entering the event site, the density increases and the dynamic zoom approach automatically applies the SpaceCont model to simulate the entrance area (second screenshot). As the simulation progresses, the zoom area slightly changes its focus in the third screenshot. In the fourth screenshot, the focus has switched from the entrance area to the event site itself.

Our conducted case study illustrates the feasibility of the developed approach. The described procedure can be even extended to macroscopic simulation models, as we discussed in previous publications \cite{Biedermann:2016:MobilTUM, Biedermann:2016:PED16}.

\subsection{Saving of computational time}
\label{cdtfasd4o4}

Spatial-continuous pedestrian dynamics approaches like the Social-Force model \cite{helbing1995social,helbing2000simulating} have a computational complexity of $\mathbf{O}(n^2\cdot m)$ with $n$ as the number of pedestrians and $m$ as the number of static obstacles in the scenario.
 Contrary, typical SpaceDisc models have a complexity of  $\mathbf{O}(n)$ with $n$ referring to the number of pedestrians in the simulated scenario \cite{Biedermann:2016:MobilTUM}. Consequently, SpaceDisc models can be calculated much faster than their spatial-continuous counterparts.

\begin{figure}
\centering
\includegraphics[width=1\linewidth]{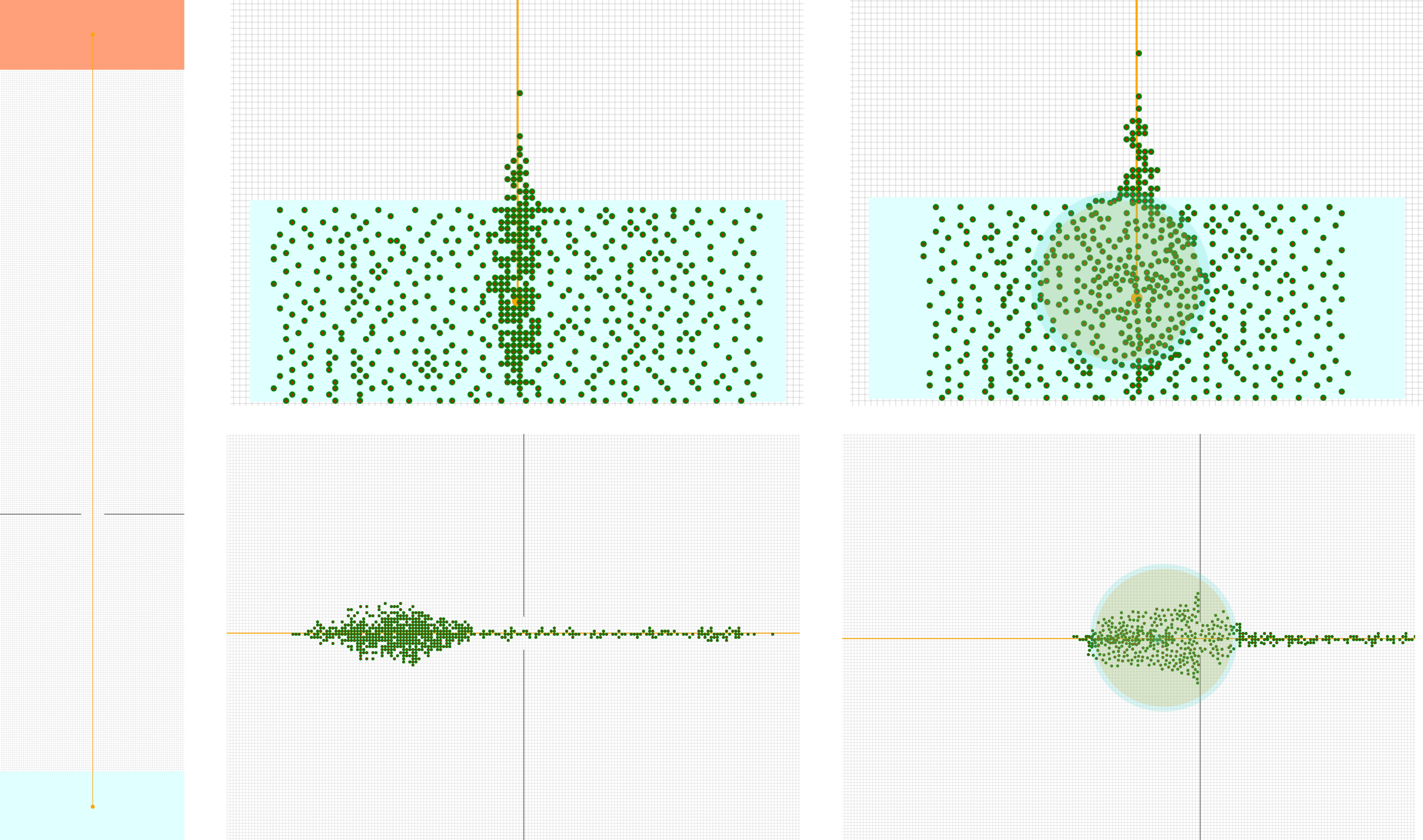}
\caption{Exemplary screenshots from the simulation study at the different time steps ($\SI{0}{\second}$, $\SI{8}{\second}$, $\SI{9}{\second}$, $\SI{70}{\second}$, $\SI{90}{\second}$) steps for 600 pedestrians in the second simulation series.}
\label{fig:7}
\end{figure}

However, the saved computational effort under real circumstances can be quite different since some computational effort has to be used for the transformation process between the coupled pedestrian dynamics models. Therefore, we made a study under real conditions to determine the efficiency of our proposed generic transformation approach. We simulated the same artificial scenario layout with three different simulation approaches and compared the needed computational time. The scenario itself can be seen on the left side of Figure \ref{fig:7}. We used the same strategical and tactical approach for all three simulation models to enable a comparability of the computational time of each model. As a strategical layer a simple origin-destination model was chosen since only one origin and one destination exists. The shortest path approach was used as the tactical layer. Each simulation was executed until every pedestrian has left the scenario. The operational model was implemented via a Social-Force model \cite{helbing1995social} for the spatial-continuous approach and via the Stock-Model for the spatial-discrete approach \cite{Biedermann:2016:MobilTUM}.

We compared the computational time for the singular simulations from 0 to 1000 pedestrians in steps of 10 pedestrians. Three different simulation series were executed. The current local densities and spatial-continuous areas were calculated automatically every $\SI{2.5}{\second}$.  The first one was simulated with the SpaceCont model only as their operational layer. The second and third simulation series used the \TransiTUM{} approach and used the SpaceDisc and SpaceCont models depending on the local density. However, different boundary conditions were used for the second and third simulation series. The second series transformed to the SpaceCont model if the density extended $\SI{1.5}{ped\per\meter\tothe{2}}$ in an area with a radius of $\SI{3.0}{\meter}$. Contrary, the third simulation series transformed to the SpaceCont model only if the density gets larger than $\SI{4.0}{ped\per\meter\tothe{2}}$ in an area with a radius of $\SI{2.0}{\meter}$.
\begin{figure}
\centering
\includegraphics[width=1\linewidth]{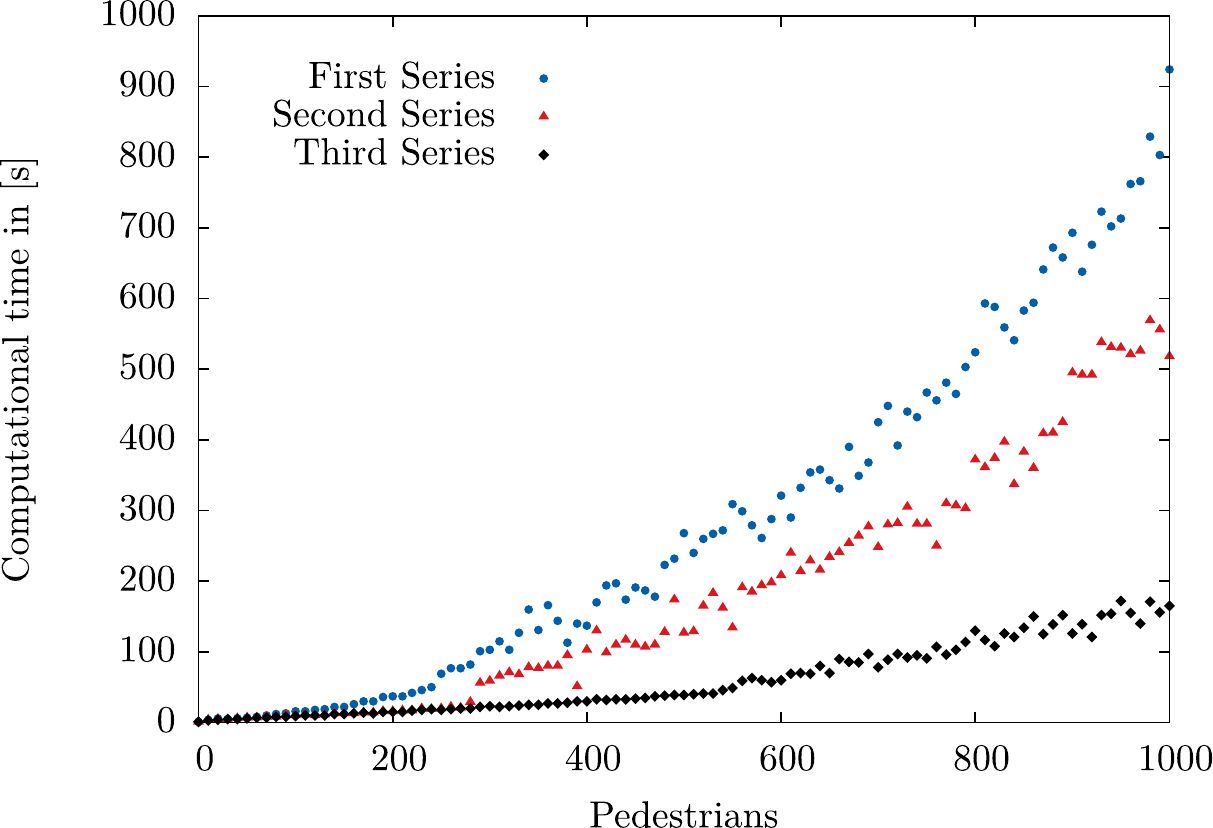}
\caption{Necessary computational time for the simulation in depending on the number of simulated pedestrians}
\label{fig:8}
\end{figure}

These simulation series show the limits of our generic approach as well as boundary conditions for their successful application (see Figure \ref{fig:8}). Though, our generic approach needed less computational effort than a pure spatial-continuous simulation, the efficiency win for the second simulation series was small by reducing the computational time only by twenty to forty percent. The reason for this unsatisfying results is quite easy: since we chose a quite low tolerable density, nearly all pedestrians were transformed. Consequently, nearly the same computational effort was necessary. However, in the third series we increased the necessary tolerable density and although we lowered the minimal radius for a transformation we could decrease the computational effort to one quarter compared to the effort for a pure spatial-continuous simulation. This shows, that the \TransiTUM{} approach is most efficient if only a small part of the whole simulation scenario has to be simulated by a SpaceCont model. Furthermore, the saving of computational time depends on the relative efficiency of the spatial-discrete model compared to its spatial-continuous counterpart. The higher the relative computational effort of the SpaceCont model, the larger the computational savings that can be accomplished by the \TransiTUM{} approach.

Due to technical progress, the computational progress increased significantly over the last years and decades. However, realistic human behavior is still far to complex to be calculated. Therefore, we have to use simplified models to predict e.g. human walking behavior. With our approach it is possible to integrate high complex models in larger scale scenarios, which otherwise could only be used for scenarios with very few pedestrians. Another benefit is the applicability of pedestrian dynamic simulations to the real world. For event managers it can be helpful to simulate pedestrian behavior to detect dangerous situation on their event site in an early stage. To increase the usability, these simulations should be easily executable on a small mobile device in a short amount of time.

\begin{figure}
	\centering
	\includegraphics[width=1\linewidth]{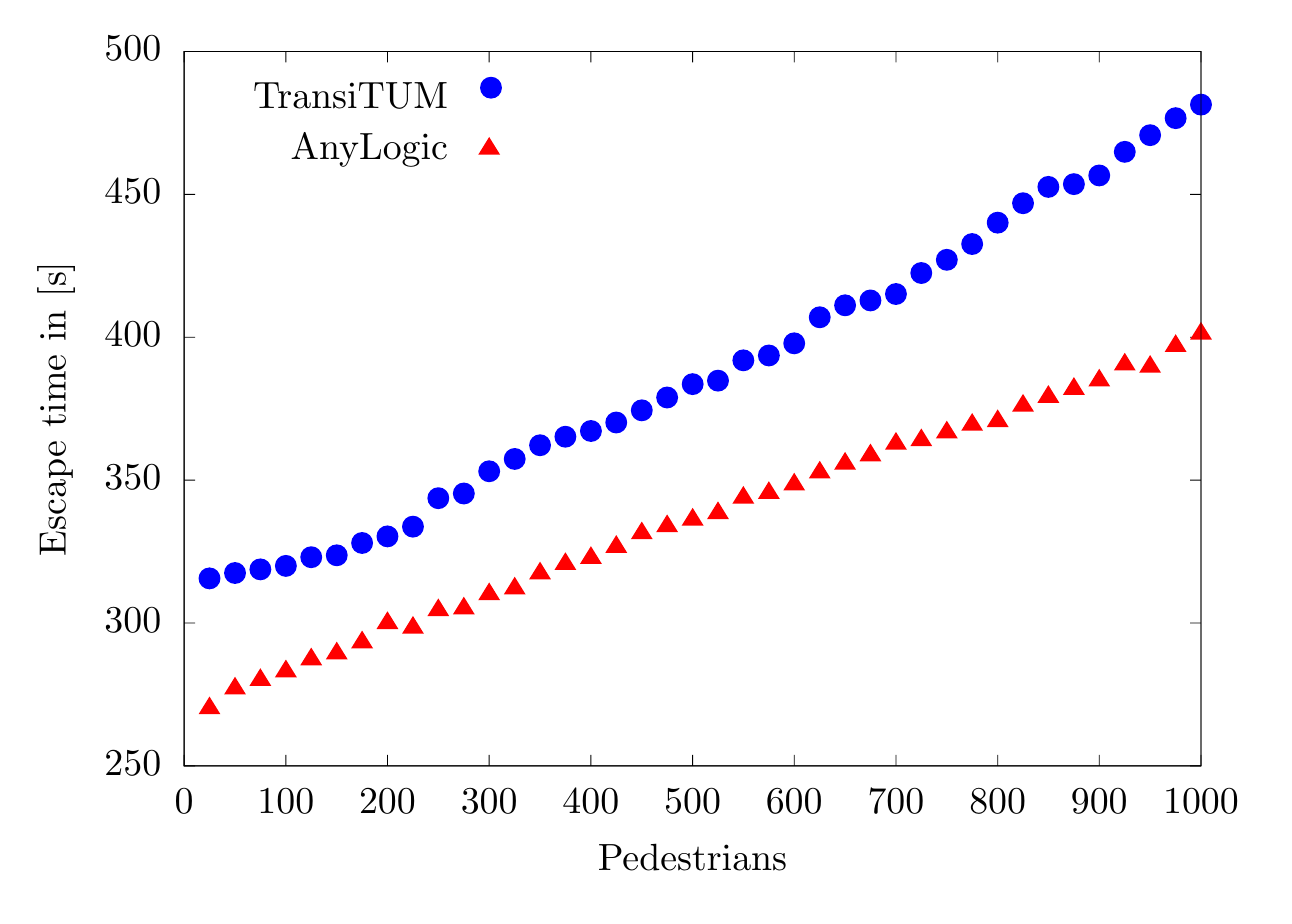}
	\caption{Comparision of escape times for the same scenario between TransiTUM and AnyLogic}
	\label{fig:escape-time}
\end{figure}

In a further research study we tested the results of our simulator in comparison to the well-established simulation software AnyLogic. We used the same scenario as for the previous simulation series. We compared the necessary escape times for the pedestrians to leave the scenario in our simulator with the escape times calculated in AnyLogic. 

Since we have to validate the whole simulation framework, we have to use non-local parameters. The global escape time is the most suitable parameter for this procedure, since the transformation framework and all underlying models contribute and influence this value. Local parameters like the flow rate at a bottleneck are only influenced by the used model at this area. Therefore, a validation of such parameters would only validate the local model, but not the whole simulation framework.

Each simulation was executed three times for each simulator and the averaged results are shown in Figure \ref{fig:escape-time}. The results were in good accordance with the results of the established AnyLogic simulator. We tested for up to 1000 pedestrians. However, due to the different used simulation models, larger differences may occur for larger pedestrian numbers.


\section{Conclusion and outlook}
\label{OutlookandConclusionas}

We introduced a new multiscale approach which is able to combine arbitrary spatial-continuous and spatial-discrete pedestrian dynamics models. Such a generic method as well as the automatic, density-depending zoom-approach had not been investigated in the field of pedestrian dynamics before \cite{Ijaz2015}. The transition between different spatial scales is executed via transit areas, which surround all spatial-continuous regions of the simulated scenario. If a pedestrian enters a transit area, the \TransiTUM{} approach detects if the pedestrian is heading to the region of another spatial scale. Pedestrians, which are heading to regions of another spatial scale, are transformed into the corresponding simulation model. An automatic zoom-approach was developed, since critical areas of the simulation scenario can change during the simulation run. The zoom-approach detects regions with high densities and switches the simulation model of these regions from the SpaceDisc model to the SpaceCont model. If the density of this region gets uncritical again, the pedestrians are transformed back to the spatial-discrete simulation model. On that way, a reduction of simulation time is possible without losing accuracy in the critical parts of the simulation.

Some limitations exist for the use of our generic transition approach. Due to its generic nature, a transformation of model specific attributes (e.g. group behavior) is not possible. Furthermore, each coupled simulation model has to fulfill some boundary conditions to ensure a smooth transformation procedure: the time steps of the spatial-discrete model have to be larger than the spatial-continuous one, each cell can contain a maximum of one pedestrian and the lattice has to consist of regular, convex grid cells. However, these conditions were fulfilled by all models we analyzed (see Table \ref{tab:MiMeTiSt}). Another issue is that the transformation can not be executed if the transit area gets overcrowded. However, the dynamic zoom approach helps to solve this problem.

Some improvements are planned for future research. We plan to include an internal parameter-exchanger to enable the \TransiTUM{} framework to include the transformation of model specific attributes. The parameter-exchanger is an internal database of the transition framework, which contains the global identification numbers of the pedestrians and their model specific attributes. On that way, the simulation framework keeps its generic nature, but is still able to transform model specific parameters. Another improvement would be 
the inclusion of scenario specific density information. For example the entry and exit areas can be given a lower maximal density limit than e.g. the dance floor. This ensures that critical areas are simulated in a sufficient high spatial resolution.

\section*{Acknowledgments}
We would like to thank Peter M. Kielar and Michael Seitz for fruitful discussions regarding the paper's subject. Additionally, we would like to thank Alexander Braun and Peter M. Kielar for the creation of the CAD Layout of the simulation scenario. Furthermore, we extend special thanks to all other contributors of the \MomenTUM{} simulator. This work was supported by the Federal Ministry for Education and Research
(Bundesministerium  f\"{u}r  Bildung  und  Forschung,  BMBF),  project  MultikOSi,
under grant FKZ 13N12823.


\bibliography{mybibfile}

\end{document}